\documentclass{mn2e}
\usepackage{subfigure}
\usepackage{epsfig}
\usepackage{amsmath}
\usepackage{graphicx}

\newcommand{\msun}{M_{\odot}}
\newcommand{\ion}[2]{#1$\;${\small\rmfamily{#2}}\relax}

\title[Galaxy Evolution:  Pair Interactions versus Environment]{Effects on Galaxy Evolution:  Pair Interactions versus Environment}
\author[S. Tonnesen and R. Cen]{Stephanie Tonnesen$^{1}$\thanks{E-mail:  stonnes@astro.princeton.edu (ST);  cen@astro.princeton.edu (RC)} and Renyue Cen$^{1}$\\
$^{1}$Department of Astrophysics, Princeton University, Peyton Hall, Princeton, NJ, 08544\\
}
\begin{document}

\pagerange{\pageref{firstpage}--\pageref{lastpage}} \pubyear{2011}

\maketitle

\label{firstpage}

\begin{abstract}

In a hierarchical universe, mergers may be an important mechanism not only in increasing the mass of galaxies but 
also in driving the colour and morphological evolution of galaxies.  We use a large sample of $\sim1000$ 
simulated galaxies of stellar mass greater than $10^{9.6}\msun$ (for $\sim4800$ observations at multiple redshifts) from  
a high-resolution (0.46 $h^{-1}$ kpc) cosmological simulation 
to determine under what circumstances being a member of a pair influences galaxy properties at z $\leq$ 0.2.  We identify gravitationally bound pairs, and find a relative fraction of blue-blue (wet), red-red (dry), and blue-red (mixed) pairs that agrees with observations (Lin et al. 2010).  All pairs tend to avoid the extreme environments of clusters and void centres.  While pairs in groups can include galaxies that are both blue, both red, or one of each colour, in the field it is extraordinarily rare for pair galaxies to both be red.
We find that physically bound pairs closer than 250 $h^{-1}$ kpc tend to have higher sSFRs than the galaxy population as a whole.  However, the sSFR of a bound galaxy relative to galaxies in a comparable local density environment (determined by the distance to the fifth nearest neighbor, $\rho_5$), differs depending on the local density. 
In regions of high $\rho_5$ the bound population has a higher fraction of star-forming (bluer) galaxies, whereas there is very little difference between bound and unbound galaxies in low $\rho_5$ regions.  This effect on the star-forming fraction may be driven by the higher fraction of bound \ion{H}{I}-rich galaxies compared to unbound galaxies, particularly at high local densities.
It appears that being in a pair has an incremental, but not overwhelming, effect on the star formation rate of the paired galaxies,
compared to the more pronounced trend where galaxies overall have low sSFR (are red) in clusters and higher sSFR (blue) at the centre of voids.  This trend depends most strongly on local galaxy density ($\rho_5$).
We find no strong evidence that pair interactions are the driver of the colour-density relation for galaxies. 

\end{abstract}

\begin{keywords}
galaxies: clusters, galaxies: interactions, methods: N-body simulations
\end{keywords}

\section{Introduction}

Observations of galaxy populations at different redshifts indicate that the red sequence has roughly doubled in mass from z $\sim$ 1 to 0, while the mass in the blue cloud remains relatively unchanged (Bell et al. 2004; Willmer et al. 2006; Faber et al. 2007; Martin et al. 2007; Ruhland et al. 2009).  Gravitational interactions between galaxies can influence their star formation rate (SFR), gas content, and morphology.  As such, they can increase the mass of star-forming galaxies, and be an important mechanism causing the evolution of blue, star-forming galaxies into red and dead galaxies.  Slow interactions that result in mergers will have a stronger impact on galaxy properties (but see Moore et al. 1996 for a discussion of the possible impact of many quick galaxy interactions).  Dry mergers between gasless galaxies can also cause the hierarchical build-up of massive red galaxies.

Galaxy-galaxy interactions may drive the evolution of blue galaxies into red galaxies by causing a burst of star formation that may exhaust the gas in the merging pair and result in a more massive red galaxy.  Larson \& Tinsley (1978) identified interacting galaxies as peculiar galaxies, and determined that many of these galaxies had recently undergone a burst of star formation.  Shortly thereafter, a number of observers found bursts of star formation associated with tidal interactions (e.g. Condon et al. 1982; Keel et al. 1985; Kennicutt et al. 1987).  Since then there has been much more observational work verifying and expanding on these results.  For example, Park et al. (2008) find that tidal effects can accelerate the consumption of gas in galaxies and transform late types into early types.  In addition, the level of star formation rate enhancement increases for closer pairs (e.g. Alonso et al. 2004, 2006; Ellison et al. 2008), more gas-rich pairs (Nikolic et al. 2004; Woods \& Geller 2007), and more evenly matched mass-ratio pairs (Woods \& Geller 2007; Ellison et al. 2008; but also see Nikolic et al. 2004).  
  
In addition to an increase in the total mass of red systems over time, there is also evidence of an environmental dependence of the red to blue galaxy ratio.  Since about z $\sim$ 0.4 red galaxies have dominated the galaxy populations of dense regions, such as clusters (Butcher \& Oemler 1978).  Cooper et al. (2008), using the DEEP2 Galaxy Redshift Survey, find that the red fraction is higher in regions of high density to beyond a redshift of 1.  If galaxy mergers are the main way to quench star formation and turn galaxies from blue to red, then it might be expected that pairs would be found in or near groups and clusters in order to match the morphology-density and colour-density relations (Oemler 1974; Dressler 1980).  Lin et al. (2010) examine the local densities in which wet, dry and mixed mergers occur and find that they broadly agree with the predictions of the colour-density relation.  Using simulations, Barton et al. (2007) find that a higher fraction of close pairs than of the unpaired galaxy population lie in dark matter halos containing more than two galaxies.  If these pairs merge, then this supports the idea that mergers are an important process driving the colour-density relation.

Although in this paper we will only focus on the interactions between bound galaxies, it is important to note that the colour-density and morphology-density relations may also be caused by a number of other possible interactions.  In addition to pair interactions, galaxies in groups and clusters can undergo a number of fast interactions with other satellite galaxies, resulting in galaxy harassment (Moore et al. 1996).  A galaxy can also interact with the cluster potential, which can strip material from the galaxy or induce star formation (Merritt 1984; Byrd \& Valtonen 1990).  Finally, a galaxy can interact directly with the intracluster medium, which can remove gas through ram pressure stripping (Gunn \& Gott 1972), thermal evaporation (Cowie \& Songaila 1977), Kelvin-Helmholtz instabilities (Chandrasekhar 1961), or viscous stripping (Nulsen 1982).  Many of these processes could also only cause starvation, or the removal of halo gas, which will slowly redden a galaxy as it exhausts its gas reservoir through star formation (Larson, Tinsley \& Caldwell 1980). 

While the mechanisms described above depend on a dense environment to affect individual galaxies, the opposite may be true for mergers.  The high velocity dispersion in clusters makes them unfriendly environments for pairs (e.g. Ostriker 1980; Makino \& Hut 1997), and it has been proposed that ``pre-processing" of galaxies before they enter clusters can cause the morphology-density relation (Zabludoff \& Mulchaey 1998; Zabludoff 2002; Mihos 2004).  For example, galaxies in groups and field pairs can undergo slower encounters that are more likely to result in mergers.  Moss (2006) observe galaxies in 8 low-redshift clusters, and find that 50-70\% of the infall population was interacting or showed a disturbed morphology indicating a recent merger.  This agrees with the results of McGee et al. (2009) and DeLucia et al. (2011), who use semi-analytic models and the Millennium Simulation (Springel et al. 2005) to find that 40-60\% of galaxies with stellar masses above 10$^9$ M$_{\odot}$ enter clusters with no companions.  Berrier et al. (2009), using a somewhat different method, examine 53 clusters in dark-matter only N-body simulations and find that 70\% of galaxies with total masses above 10$^{11.5}$ M$_{\odot}$ fall into clusters with no companions.

There continues to be a large amount of observational work on how interactions affect the SFR of galaxies in a range of environments.   Wong et al. (2011) measure star formation rate enhancements of 15-20\% in isolated pairs, and also find marginal evidence for increasing enhancement with decreasing redshift.    Barton et al. (2007) examine isolated close pairs and find that SFR can be boosted by a factor of 30.  They argue that studies of pairs in denser environments are flawed because local galaxy density does not accurately determine whether galaxies are in their own halo or are satellite galaxies.  Despite this concern, there has been significant observational work on how the larger-scale environment can affect the SFR of interacting galaxies using local galaxy density.  Lambas et al. (2003) and Alonso et al. (2004) determine that pairs need to be closer in groups than in the field for the interaction to trigger enhanced star formation.  Ellison et al. (2010) examine whether pairs behave differently in different local density environments, which they determine by using the projected distances to the fourth and fifth nearest neighbors.  They find that in low density environments (log $\Sigma <$ -0.55), interacting pairs trigger star formation, while pairs in high-density environments do not have enhanced star formation (log $\Sigma >$ 0.15).  Using SDSS data, Perez et al. (2009) compare pair galaxies to control samples that were matched for dark matter halo mass, stellar mass, redshift, and local galaxy density (within the distance to the fifth nearest neighbor, $\Sigma_5$).  They find that although star formation is higher in pairs in low local density regions than in high local density regions, it is more enhanced in pairs in mid-range density regions relative to unpaired galaxies.  They suggest that this is evidence for galaxy pre-processing in intermediate-density regions before entering clusters. 

Patton et al. (2011) use the Sloan Digital Sky Survey Data Release 7 to identify more than 20,000 pairs and compare them with control galaxies matched in stellar mass and redshift.  They find a higher red fraction in paired galaxies, which they determine is likely due to the higher density environments in which they are found.  They also find a population of extremely blue close pairs, indicating interaction-induced star formation in their paired galaxies (also observed by Alonso et al. 2006; Perez et al. 2009; Darg et al. 2010). 
 
Perez et al. (2006a,b) use GADGET-2 to study the star formation, colours, and chemical abundances of galaxies in pairs.  When comparing their pairs to their control sample, they find strong agreement with observational trends:  pairs are bluer than non-paired galaxies, which is caused by their higher star formation rate.  Also, closer pairs are bluer with higher SFRs.  They find that both pairs and control galaxies follow the observed colour-density relation, but that the trends are stronger for close pairs.  Specifically, Perez et al. (2006a) find a higher fraction of active galaxies in pairs in all environments, but the largest difference when comparing the fraction of actively star-forming galaxies is in low-$\Sigma$ regions ($\Sigma <$ 0.8 Mpc$^{-2}$).  They claim that this indicates that galaxy-galaxy interactions help establish the colour-density relation.  They also find that the stellar populations in paired galaxies are more metal-enriched than non-paired galaxies (Perez et al. 2006b).  This is in contrast to the observations by Kewley et al. (2006; 2010), who find lower metallicities in interacting galaxies, possibly due to gas inflow from the outer parts of the disc or halo (e.g. Hibbard \& van Gorkom 1996; Rampazzo et al. 2005).  

In this paper we use a cosmological simulation to examine the role of pairs and tidal interactions on galaxy colour, SFR, and gas mass.  We focus on two regions within a simulation with box side of 120 $h^{-1}$ Mpc, one centred around a cluster with size 21 $\times$  24 $\times$ 20 $h^{-3}$ Mpc$^3$ and the other centred around a void with size 31 $\times$ 31 $\times$ 35 $h^{-3}$ Mpc$^3$.  This gives us a broad range of local environments to study as well as two extreme large-scale environments.  We consider the galaxy populations at five redshifts:  0.0, 0.05, 0.1, 0.15, and 0.2.  This increases our sample size, makes sure that any anomalies at a single redshift snapshot do not dominate the results, and checks to see if there is evolution in the pair population across these later redshifts.  Choosing this redshift range also allows for better comparisons with recent observational work on this topic using SDSS (e.g. Patton et al. 2011).  Also, when we determine distances between galaxies, and use a distance criteria to determine the local galaxy density, we use the physical distance rather than the comoving distance.  Our goal in this paper is to determine under what conditions being a member of a pair or group affects galaxy properties.  We examine this question in the context of the larger scale environment of the galaxies.

After an introduction to our simulation (Section \ref{sec:boxes}), we discuss our galaxy selection technique and our method for determining observable quantities for each galaxy in Section \ref{sec:galaxyselection}.  In Section \ref{sec:projpairs} we examine the reliability of observational techniques used to identify pairs.  We then explain our method for finding gravitationally bound pairs in our simulation (Section \ref{sec:bounddef}).  In Section \ref{sec:pairdemographics} we describe the demographics of bound galaxies, including where they are in relation to the cluster or void and their local density (\S \ref{sec:wherepair}).  We also consider the colour of galaxies in pairs (\S \ref{sec:wdm}).  We then determine whether bound galaxies are different than the general galaxy population in terms of their sSFR and gas mass in Section \ref{sec:comparisons}.  After attempting to determine whether pair interactions have an effect on galaxy properties over and above any differences caused by the local density (Section \ref{sec:env}), we check whether pair separation affects our results in Section \ref{sec:interacting}.  In Section \ref{sec:story}, we synthesize our findings into a coherent physical scenario.  Finally, we summarize our main conclusions and make some additional comparisons with observations and predictions (\S \ref{sec:conclusion}).  

\section{Methodology}

\subsection{Simulation Details}\label{sec:boxes}

For details of our simulations, we refer the reader to Cen (2010), although for completeness we reiterate the main points here.  We perform cosmological simulations with the adaptive mesh refinement (AMR) Eulerian hydrodynamical code \textit{Enzo} (Bryan 1999; O'Shea et al. 2004; Joung et al. 2009).  We use cosmological parameters consistent with the WMAP7-normalized LCDM model (Komatsu et al. 2011):  $\Omega_M$ = 0.28, $\Omega_b$ = 0.046, $\Omega_\Lambda$ = 0.72, $\sigma_8$ = 0.82, $H_o$ = 100 $h$ km s$^{-1}$ Mpc$^{-1}$ = 70 km s$^{-1}$ Mpc$^{-1}$, and $n$ = 0.96.  We first ran a low resolution simulation with a periodic box of 120 $h^{-1}$ Mpc on a side, and identified two regions:  one centred on a cluster and one centred on a void at $z = 0$.  We then resimulated each of the two regions separately with high resolution, but embedded within the outer 120 $h^{-1}$ Mpc box to properly take into account large-scale tidal field effects and appropriate fluxes of matter, energy and momentum across the boundaries of the refined region.

The cluster refined region, or C box, is 21 $\times$ 24 $\times$ 20 $h^{-3}$ Mpc$^3$.  The central cluster is $\sim$2 $\times$ 10$^{14}$ M$_\odot$ with a virial radius (r$_{200}$) of 1.3 $h^{-1}$ Mpc.  The void refined region, or V box, is somewhat larger, at 31 $\times$ 31 $\times$ 35 $h^{-3}$ Mpc$^3$.  Although we name these two regions based on whether they contain the cluster or void, we emphasize that these high-resolution boxes are much larger than the cluster or the void at their centres.  Thus, there are galaxies at a range of local densities in both boxes, and there is substantial overlap of local densities between the two volumes.

In both refined boxes, the minimum cell size is 0.46 $h^{-1}$ kpc, using 11 refinement levels at $z = 0$.  The initial conditions for the refined regions have a mean interparticle separation of 117 $h^{-1}$ kpc comoving, and a dark matter particle mass of 1.07 $\times$ 10$^8$ $h^{-1}$ M$_\odot$.   The simulations include a metagalactic UV background (Haardt \& Madau 1996), a model for shielding of UV radiation by neutral hydrogen (Cen et al. 2005), and metallicity-dependent radiative cooling (Cen et al. 1995).  The fraction and density of neutral hydrogen is directly computed within the simulation.  The computed \ion{H}{I} gas roughly corresponds to all cold gas with temperature less than about 2 $\times$ 10$^4$ K, above which
collisional ionization becomes important, causing the ionized hydrogen to be the dominant ionization state.  Star particles are created in cells that satisfy a set of criteria for star formation proposed by Cen \& Ostriker (1992), and supernovae feedback is included (Cen et al. 2005).  Each star particle has a mass of $\sim$10$^6$ M$_\odot$, which is similar to the mass of a coeval globular cluster.

\subsection{The Galaxy Sample} 

\subsubsection{Galaxy Sample Selection}\label{sec:galaxyselection}

We use HOP (Eisenstein \& Hut 1998) to identify galaxies using the stellar particles.  This has been tested and is robust using reasonable ranges of values (e.g. Tonnesen, Bryan, \& van Gorkom 2007).  In order to choose only well-resolved galaxies, we only consider those galaxies with a stellar mass greater than 10$^{9.6}$ M$_\odot$.  In the C box, this leaves 61\% of the originally identified galaxies, and in the V box this leaves 49\% of the galaxies.  We also demand that all of the dark matter particles in each galaxy be highly refined--so each galaxy must have resided in the refined region since the beginning of the simulation.  In Table \ref{tbl-pairs}, we show the number of galaxies above our minimum mass at each output in each box, and see that this number tends to decrease with decreasing redshift.  While the fraction of galaxies with M$_*$ $>$ 10$^{9.6}$ M$_\odot$ is highest in the z = 0 boxes, the number of galaxies that began in the refined region and have stayed in the refined region has decreased.  In the C box, this is due to both mergers and galaxies leaving the refined region, while the underdensity of the V box results in galaxies leaving the refined region.

We plotted projections of the star particles of each of these galaxies in order to verify first that HOP was identifying galaxy-like objects (with a density peak), and second, that HOP was not grouping multiple galaxies together.  In both the Cluster and Void boxes, only a few of the galaxies above our minimum mass that HOP identified had density profiles without a strong density peak, eliminating 8 galaxies in the C box and 7 galaxies in the V box over all the redshift outputs.  In addition, in the C box 47 HOP-identified galaxies in fact had two density peaks, and one had three ``galaxies".  In the Void box a total of 13 HOP-identified galaxies had two density peaks.  Both of these problems add up to a misidentification of only 2\% in each box.  

If we treat multiple density peaks in HOP-identified galaxies as multiple galaxies, we have 4858 galaxies above the minimum mass constituting our ``total" galaxy population over all redshift outputs (C box z = 0.0, 0.05, 0.1, 0.15, and 0.2 for a total C box galaxy population of 3324; V box z = 0.0, 0.05, 0.15, and 0.2 for a total V box population of 1534). 

\subsubsection{HOP multi-peak sample}\label{sec:HOPgals}

All of the HOP-identified galaxies with multiple density peaks make up our HOP multi-peak sample (HOP has combined multiple galaxies into one).  We will discuss each of the galaxy pairs from the HOP multi-peak sample as if observers would identify them as mid-merger or strongly interacting pairs (we do not include the largest cluster cD galaxy in any of our pair samples, including the HOP multi-peak sample).  These galaxies are likely to be observationally identified as interacting for the same reason that HOP identifies them as a single galaxy -- they have either a small gap between or overlapping stellar populations, and frequently have tidal streams connecting the galaxies.  The high-density stellar cores of the galaxies are within about 30 kpc of one another.  We also checked the velocities of these galaxies and find that they are gravitationally bound and not simply close passes.   

\begin{figure}
\includegraphics{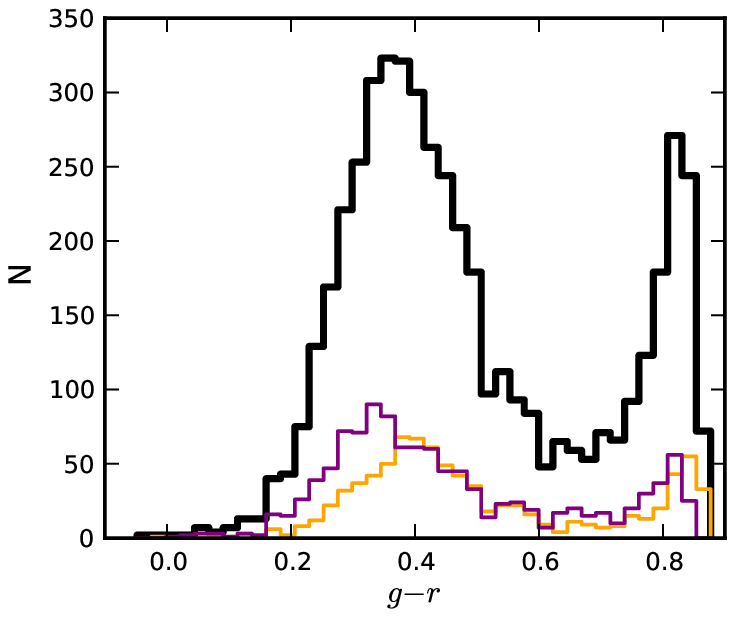}\\
\includegraphics{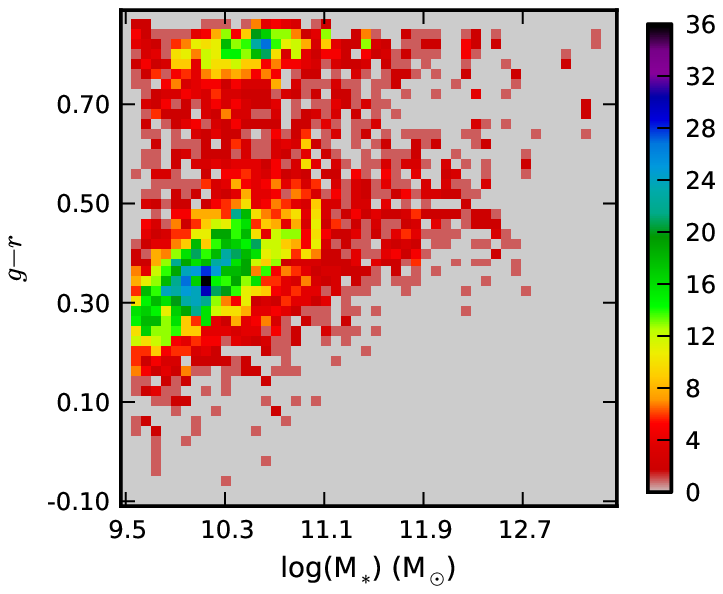}\\
\caption{Top Panel:  A histogram of the colour distribution of all the galaxies in both the C and V boxes above our minimum mass.  The distribution is bimodal, and has more blue galaxies than red galaxies.  Part of this may be explained by comparing the galaxies at z = 0.2 (purple line) to galaxies at z=0 (orange line).  The galaxies at earlier times are more plentiful and bluer.  The split between the blue and red population happens at around the same colour as in observations:  $g-r$ = 0.65.  Bottom Panel:  A two-dimensional histogram of galaxy colour and mass.  The bimodal distribution is also seen in this panel, and we see that high-mass galaxies tend to be redder than low-mass galaxies.  See Section \ref{sec:galaxyselection} for discussion.}
\label{fig-colourhist}
\end{figure}

\subsubsection{Galaxy Characteristics}

For details on the properties of our galaxy sample, see Cen (2011).  In particular, Cen (2011) compares the evolution of the SFR density from the C and V runs to observations and finds strong agreement.  Cen (2011) also compares the SDSS r-band galaxy luminosity function at z = 0 to observations and finds excellent agreement at M$_r$ $>$ -22.  We do overproduce large galaxies, and including AGN feedback can bring the luminosity function into agreement even at high luminosities (as shown in Figure 3 of Cen (2011)).  

In the top panel of Figure \ref{fig-colourhist}, we plot the colour distribution of our total galaxy sample (recall our ``total galaxy sample" is composed of all galaxies with M$_{star}$ $>$ 10$^{9.6}$ M$_\odot$).  We see that there is a bimodal distribution of colours as in observations, and the total range of $g-r$ colours in our simulations is similar to the observed range of galaxy colours (e.g. Blanton et al. 2003; Patton et al. 2011).  However, the blue peak is slightly bluer (about 0.05-0.1 magnitudes) and the red peak is about 0.1 magnitudes bluer than in Blanton et al. (2003), and our distribution is narrower (particulary on the red end of the distribution).  

In the bottom panel of Figure \ref{fig-colourhist}, we plot the two-dimensional histogram of galaxy colour and mass.  Again, we see the bimodal distribution of the panel above, and we also see that our bluest galaxies tend to be low-mass, and our high-mass galaxies are redder than low-mass galaxies, in broad agreement with observations (Baldry et al. 2004).  This mass-colour relation has begun to be understood in a cosmological context by considering the gas properties of galaxy halos:  high stellar mass galaxies have higher mass halos containing hot gas that either cools slowly before it can form stars, or cannot radiatively cool to form stars, while galaxies in low-mass halos are able to directly accrete cold gas that can quickly form stars (e.g. Dekel \& Birnboim 2006; Kere{\v s} et al. 2005).  The ratio of the number of galaxies in the blue cloud to that on the red sequence depends on the uncertain weightings for the C and V boxes, making a direct comparison to observations difficult.  In the bottom panel of Figure \ref{fig-colourhist} we see that there are a few unrealistically large galaxies--these are mainly at the centre of our cluster or at the centres of groups.  The inclusion of AGN feedback, as mentioned above and in Cen (2011), would largely mitigate this discrepancy between our simulations and observations.  

We can partially explain the differences between the colour distribution of our sample from z = 0 to z = 0.2 to the Blanton et al. (2003) observations (at z = 0.1) by comparing the histograms from the z = 0 (orange line) and z = 0.2 (purple line) outputs.  The z = 0.2 output is bluer at both peaks, and contains more galaxies (see also Table \ref{tbl-pairs}).  We expect our (large) galaxies to become more red with time, as cold gas can no longer be accreted by galaxies as they grow or enter larger halos or overdense large-scale structure (Cen 2011).  The fact that both blue and red galaxies were bluer in the past has been observed (e.g. Blanton 2006, although for a larger redshift range).  In addition, 
these differences may be in part due to the fact that we do not include dust reddening for our galaxies, 
in part because we do not include all of the physical 
processes that could affect the colours of these galaxies (such as feedback from AGN and SN Ia),
and perhaps in a large part because our total galaxy sample (across both the C and V boxes) is not necessarily equal to the global average sample.
For these reasons we will focus mainly on comparative studies within our simulations, which should be much less dependent on these uncertainties. 

Kreckel et al. (2011) closely examine galaxies in the same V box as our simulation, and in fact we use the void centre point that they identified.  Their work differs from ours in that they have one fewer level of refinement (two times less resolution), and use somewhat different criteria in their HOP galaxy identification.  This results in differences of less than a factor of two in the measured properties of 
our populations.  By far the biggest difference is that they include galaxies well below our lower mass limit in their analysis.  

The luminosity of each stellar particle in each of the five Sloan Digital Sky Survey (SDSS) bands is computed using the GISSEL stellar synthesis code (Bruzual \& Charlot 2003), by supplying the formation time, metallicity and stellar mass.  Collecting luminosity and other quantities of member stellar particles, gas cells and dark matter particles yields the following physical parameters for each galaxy:  position, velocity, total mass, stellar mass, gas mass, mean formation time, mean stellar metallicity, mean gas metallicity, star formation rate, and luminosities in the five SDSS bands.

\section{Can Observers Pick out Pairs?}\label{sec:projpairs}

In this paper we will compare our results to observational trends in order to gain physical insight into what causes pairs to be different from galaxies without a bound companion.  It is worthwhile first to ask whether observers are in fact picking out real pairs since they do not have 
real-space three dimensional information.  In order to determine whether the observational selection criteria for pairs actually choose galaxies that are close to one another and/or that are gravitationally bound to one another, we choose ``projected pairs" using a few sets of criteria used in recent observational work.  The least stringent criterion for choosing projected pairs is from Perez et al. (2009), who use a projected distance (r$_p$) of less than 100 kpc $h^{-1}$ and relative line-of-sight velocities $\Delta$v $<$ 350 km s$^{-1}$.  We also use two criteria from Patton et al. (2011), who use a relative line-of-sight velocity criterion of $\Delta$v $<$ 200 km s$^{-1}$, and a r$_p$ upper limit of either 60 kpc $h^{-1}$ or 30 kpc $h^{-1}$.  We choose projected pairs using the x-y, x-z, and y-z planes in our simulation for both our C box and V box, and the velocity difference perpendicular to the plane as our line of sight velocity.  In Table \ref{tbl-projpairs} we show the number of pairs we find for each selections criterium across all projections and all redshifts. 

\begin{table*}
\begin{center}
\caption{Number of Projected Pairs\label{tbl-projpairs}}
\begin{tabular}{c | c | c | c}
\hline
Box & r$_p <$ 100 kpc $h^{-1}$ & r$_p <$ 60 kpc $h^{-1}$ & r$_p <$ 30 kpc $h^{-1}$ \\
(all projections) & $\Delta$v $<$ 350 km s$^{-1}$ & $\Delta$v $<$ 200 km s$^{-1}$ & $\Delta$v $<$ 200 km s$^{-1}$ \\
(all redshifts) & & & \\
\hline
C & 1512 & 513 & 201\\
V & 280 & 144 & 62\\
\hline

\end{tabular}
\end{center}
\end{table*}

\begin{figure}
\includegraphics{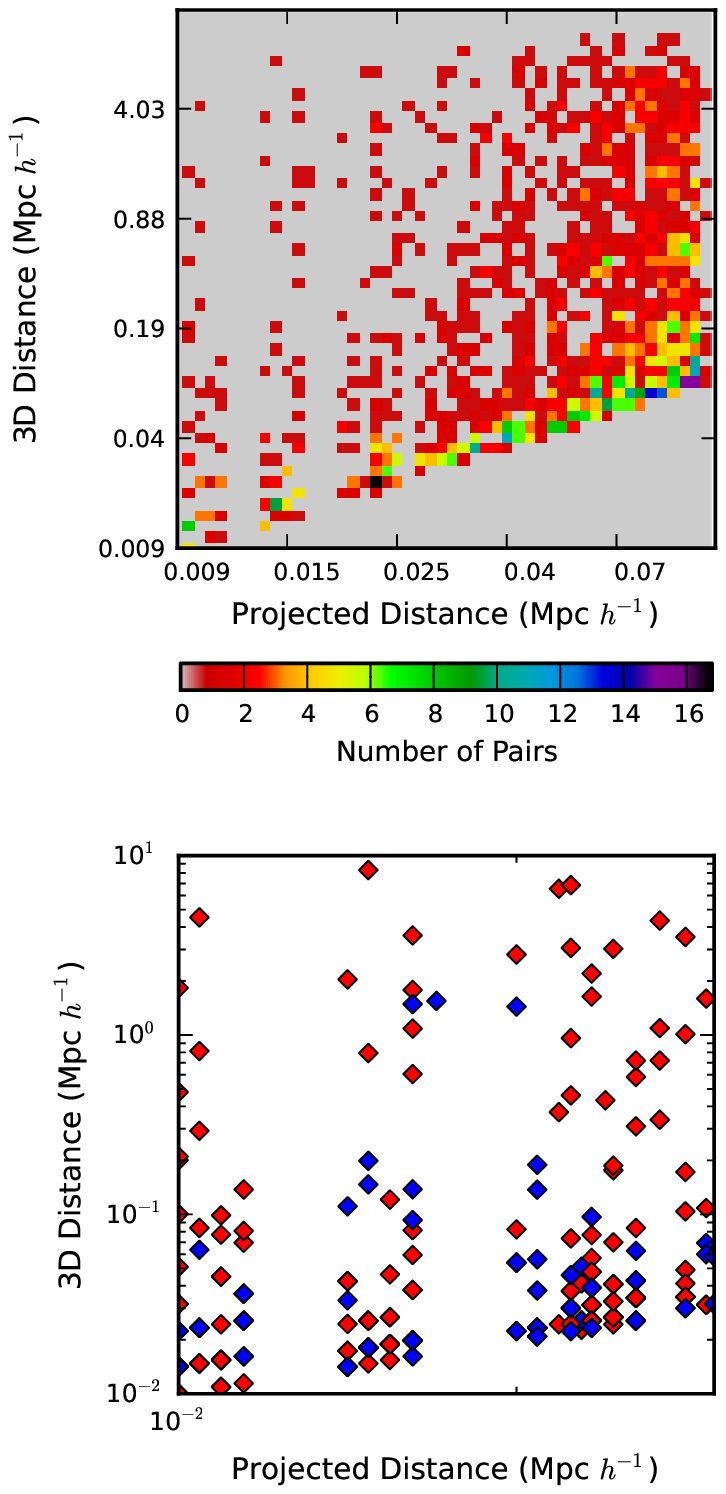}
\caption{The two-dimensional and three-dimensional distance between projected pairs selected in our simulation.  The top panel shows a two-dimensional histogram of projected pairs chosen to be within 100 kpc and 350 km s$^{-1}$.  The bottom panel shows only the pairs that are within 30 kpc and 200 km s$^{-1}$.  Blue symbols are from the V box and red are from the C box.  Clearly any of these pair-selection criteria allow for a large range of actual distance.}
\label{fig-distprojdist}
\end{figure}

We first consider whether galaxies that are close in projected distance and radial velocity are in fact close when considering all three dimensions in
real space.  As shown in Figure \ref{fig-distprojdist}, none of the three sets of criteria consistently select galaxies that are actually close to one another.  The top panel is a two-dimensional histogram of the projected pairs that are within 100 kpc and 350 km s$^{-1}$.  As larger projected distances are allowed, more projected pairs actually have large three-dimensional distances.  Sixteen of these projected pairs are more than 8 $h^{-1}$ Mpc apart (one with a projected distance of less than 20 kpc).  The projected pairs with the largest three-dimensional distances are from the C box (the largest three-dimensional distance in the V box is $\sim$ 3 $h^{-1}$ Mpc).  This is because C box projected pairs contain both galaxies within the cluster that happen to have small line-of-sight velocity differences and pairs between a cluster member and a foreground or background galaxy that have small line-of-sight velocity differences.  This highlights the well-known concern that choosing projected pairs may produce spurious pairs near clusters where there is a high density of galaxies (Mamon 1986; Alonso et al. 2004; Perez et al. 2006a), but we also find spurious pairs in our lower-density large scale environment. 

In the bottom panel we plot projected pairs found using the most stringent criterion from Patton et al. (2011):  $\Delta$v $<$ 200 km s$^{-1}$ and r$_{p} <$ 30 $h^{-1}$ kpc.  Red symbols are C galaxies, and the blue are V galaxies.  Even this most strict criterion finds galaxy  projected pairs that are separated by large three-dimensional distances.  

\begin{figure}
\includegraphics{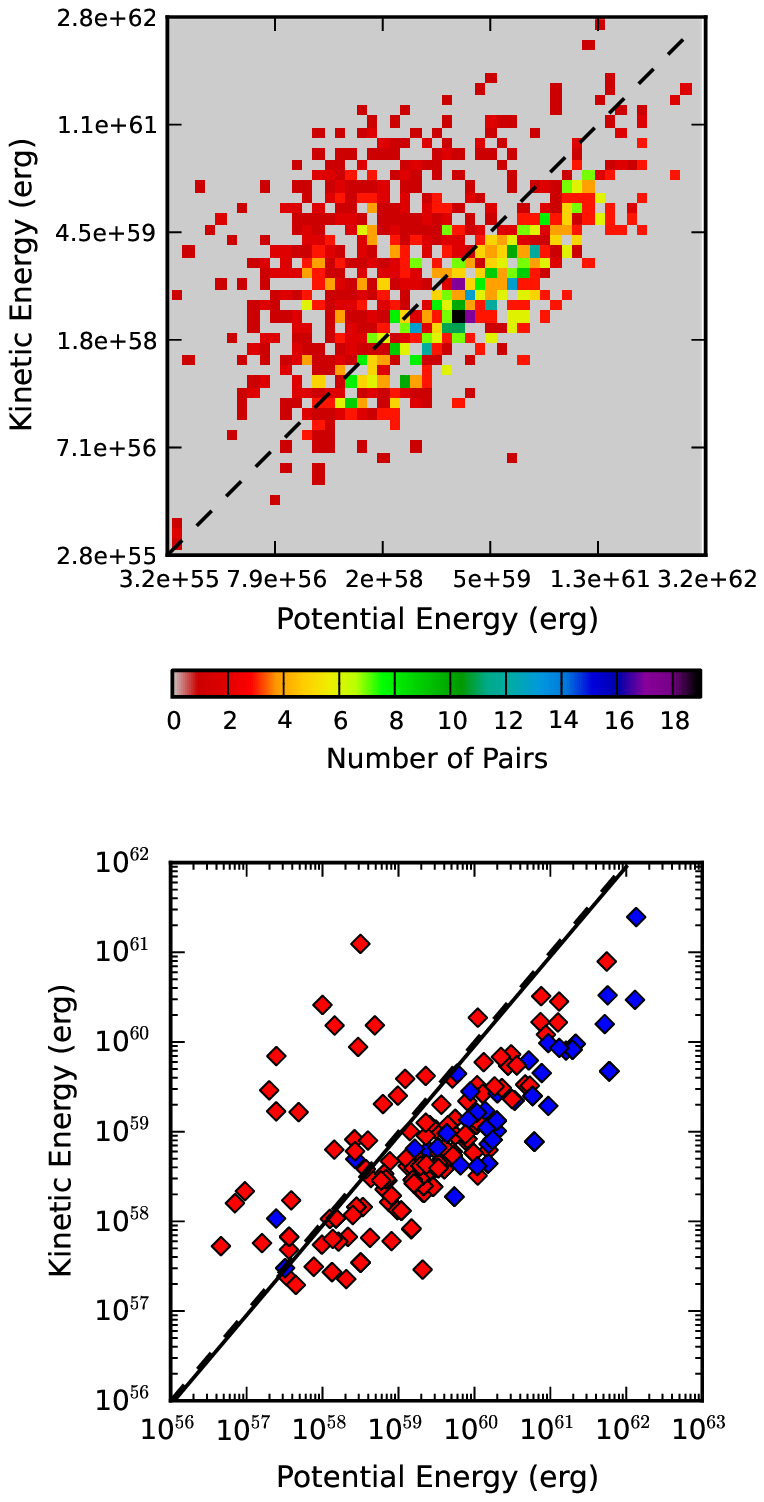}
\caption{The kinetic energy plotted against the potential energy of the pair system.  The line denotes K =0.9 $|$W$|$ (a K=$|$W$|$ line is essentially the same).  The colours and symbols are as in Figure \ref{fig-distprojdist}.  These criteria do not necessarily identify bound pairs.}
\label{fig-etotprojdist}
\end{figure}

In the rest of this paper, we will be considering only galaxies that are gravitationally bound, which we define as a pair of galaxies whose total kinetic energy in the centre-of-mass frame is less than 90\% of its potential energy.  This definition will be discussed in greater detail below (Section \ref{sec:bounddef}).  Here we calculate whether the projected pairs identify bound galaxies.  In Figure \ref{fig-etotprojdist} we plot the kinetic energy versus the potential energy.  Again, the top panel is for the least strict pair selection criterion, and the bottom panel shows each pair selected with the most strict criterion.  As a more strict selection criterion is used, more galaxies are bound ($\sim$11\% using the most strict criterion).  Also, projected pairs in the C box are more likely to have a positive total energy.

A similar analysis has been undertaken by Perez et al. (2006a).  In order to determine if projected pairs are indeed pairs, they use a three-dimensional distance criterion of 100 $h^{-1}$ kpc.  They find that using the least strict selection criterion ($\Delta$v $<$ 350 km s$^{-1}$ and r$_{p} <$ 100 $h^{-1}$ kpc), 27\% of projected pairs are spurious.  We find about 60\% of our projected pairs have three dimensional distances $>$ 100 $h^{-1}$ kpc.  Using our gravitationally bound criterion, we find that 48\% of the identified projected pairs are in fact spurious.  This large spurious fraction is dominated by the C box projected pairs, which have a spurious fraction of 55\%.  If we only include the projected pairs in the V box and the HOP multi-peak sample, we have a spurious pair fraction of only 12\%.  Again, this agrees with the idea that projected pairs are more likely to produce spurious pairs in higher-density environments (Mamon 1986; Alonso et al. 2004; Perez et al. 2006a).  Perez et al. (2006a) determine local density from the projected distance to the fifth nearest neighbor with M$_r$ $<$ -20.5 and radial velocity differences lower than 1000 km s$^{-1}$.  If we only include galaxies at low local densities we can agree with the Perez et al. (2006a) results.  However, in order to do this we must only consider regions with $\rho_5$ $\le$ 1 $h^3$ Mpc$^{-3}$, which as we will show in Section \ref{sec:env}, includes only half of the total or pair galaxy population.  Further, at these low local densities we consider only $\sim$ 44\% of our projected pairs.  The transition from blue to red galaxies seems to occur in higher density regions--therefore, when considering if being a member of a pair is important to galaxy evolution we must consider these regions, and will lose important information by simply discarding high local density regions from our analysis.  

\section{Bound Pairs}\label{sec:bounddef}

In order to find galaxy pairs in our simulation, we calculate whether a pair of galaxies is physically bound (we will stress that not all close galaxies are gravitationally bound in Section \ref{sec:unbound}).  To do this we calculate the kinetic energy of the pair using the galaxies' velocities relative to the centre of mass velocity.  We also include the velocity due to the Hubble expansion from the centre of mass of the pair in the calculation of their velocities relative to the centre of mass.  When calculating the potential energy of the pair, we also consider the positive energy due to the Hubble expansion of the matter between the galaxies.  Our potential energy equation takes this form:

\begin{equation}
W = -\frac{G M_1 M_2}{d} + \frac{3}{10} H^2r^2 \times (\frac{4}{3}\pi \rho_m r^3),
\end{equation}
\noindent where $r = 0.5 d$ and $\rho_m = \rho_{crit} \times \Omega_{m}$.  Because we demand that our potential energy be negative, this puts an upper limit on the distance between bound galaxies (that is mass-dependent).  
We consider pairs that are slightly more tightly bound than simply $|W| > K$, so we choose a limit of $0.9 |W| > K$.  Our results do not qualitatively vary even if we use an upper limit of 0.5.  This does mean that we can have widely separated galaxies that are bound under our criteria (see Figure \ref{fig-massvsdist}).  While very widely separated galaxies will not merge within a Hubble time, and are distant enough that we would expect them to show no effects from being bound, for completeness we keep them in our sample.  As we will discuss below, we include distance cuts to focus on pairs that are more likely to be interacting. 

Although we search for pairs, we also find groups--either multiple galaxies bound to the same larger galaxy, or bound galaxies  in which one is also directly bound to another, larger, galaxy.  \textit{In the rest of this paper, the term pair refers to a galaxy that is gravitationally bound to at least one other galaxy--it includes galaxies that are a member of a single pair or a member of a larger group.}  We do not include galaxies bound to the cluster cD in our count, as this would defeat our purpose of understanding when and how being a member of a pair influences galaxy properties rather than simply being inside of a cluster.  Here we define the cD to be the central dominant galaxy of our largest cluster.  We also have groups that contain a central dominant galaxy, to which we do allow galaxies to be bound.  

In Table \ref{tbl-pairs}, we show the number of galaxies that are bound to at least one other galaxy for each of our criteria (as some galaxies are bound to multiple galaxies, the number of ``pairs" is not simply half the number of bound galaxies).  We also define subsets of pairs by imposing a distance upper limit to our bound pairs (as mentioned in the Introduction, we use the physical distance between galaxies).  Note that the criteria are inclusive:  the d $<$ 0.25 Mpc $h^{-1}$ galaxies are a subset of the d $<$ 0.5 Mpc $h^{-1}$ galaxies,  which are a subset of the $\frac{K}{|W|} < 0.9$ galaxies.  We reiterate that in our pair selection, we only consider galaxies with M$_{*}$ $\ge$ 10$^{9.6}$ M$_\odot$, which constitute our ``total" galaxy population.  We will also discuss the HOP multi-peak sample, listed in the final column of Table \ref{tbl-pairs}.  This population was chosen differently than the other pair samples, and therefore is not a subset of the bound pair populations.  Of course, if a HOP multi-peak galaxy is also bound to other galaxies, both members of the HOP pair are included in that set of bound pairs.

\begin{table*}
\begin{center}
\caption{Number of Galaxies\label{tbl-pairs}}
\begin{tabular}{c | c | c | c | c | c | c}
\hline
Box & redshift & all galaxies &$\frac{K}{|W|} < 0.9$ & d$<$0.5 Mpc $h^{-1}$ & d$<$0.25 Mpc $h^{-1}$ & HOP\\
 & & & & & & multi-peak\\
\hline
C &  0 & 547 & 228 & 102 & 24 & 14\\
C & 0.05 & 680 & 352 & 234 & 155 & 14\\
C & 0.1 & 694 & 399 & 255 & 172 & 24\\
C & 0.15 & 699 & 403 & 262 & 181 & 23\\
C & 0.2 & 704 & 421 & 280 & 201 & 22\\
V & 0 & 296 &  73 & 32 & 10 & 4\\
V & 0.05 & 412 & 190 & 107 & 75 & 4\\
V & 0.15 & 418 & 206 &112 & 71 & 10\\
V & 0.2 & 408 & 210 & 112 & 70 & 8 \\
\hline
\end{tabular}
\end{center}
\end{table*}

\section{Pair Demographics}\label{sec:pairdemographics}

\subsection{Redshift Distribution}\label{sec:redshift}

A perusal of Table \ref{tbl-pairs} quickly shows that both the total number of galaxies and the number of pair galaxies (as defined, galaxies that are gravitationally bound to at least one other galaxy) increases with redshift.  In fact, the fraction of pair galaxies at any given output also increases with redshift.  This means that trends that we see when comparing the total population to the pair population may in part be attributable to trends of the total population with redshift.  When considering the total galaxy populations, we find that, on average, galaxies at higher redshift have bluer colour distributions, higher SFRs, and  reside in regions of higher local density.  This is all in good agreement with observations (e.g. Blanton et al. 2003; Blanton 2006; Martin et al. 2007).  We also find that at higher redshifts galaxies tend to have lower stellar masses, and there are fewer galaxies with very low M$_{HI}$/M$_{*}$.  These trends can be explained by continued mass growth through star formation, and continued heating of the cosmic gas.  We address this issue in more detail and find that our results are not due to the redshift distribution of our pair galaxies (Appendix \ref{sec:redshiftexplain}).  

It is also worth noting that we are examining some of the same galaxies at multiple redshifts.  Therefore it is important to consider what is simply the evolution of the same population over time, as more gas is turned into stars and galaxy masses increase, and what is due to gravitational interactions between pair galaxies.   We also address this issue by considering each redshift output individually, and discuss these results in Appendix \ref{sec:redshiftexplain}.  

\subsection {Mass ratios and Distances}\label{sec:massdist}

\begin{figure}
\includegraphics{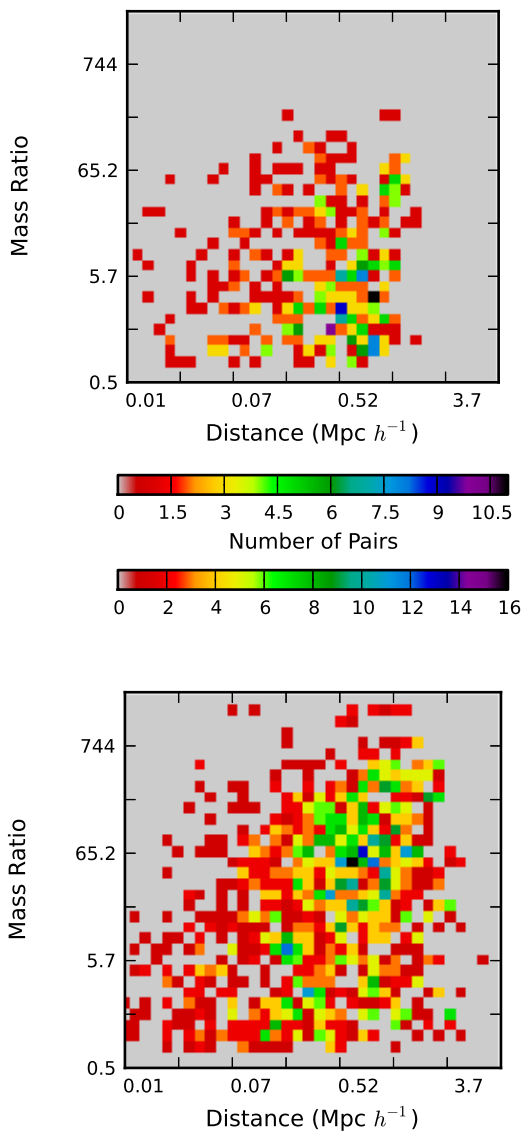}
\caption{This figure displays the mass ratio and distances between our identified pair galaxies (0.9$|$W$|$ $>$ K).  The top panel is the V box galaxies, and the bottom panel is the C box galaxies.  There is a large range in both quantities, but as we consider closer pairs the mass ratio tends to decrease.  See Section \ref{sec:massdist} for discussion.}
\label{fig-massvsdist}
\end{figure}

Observationally it has been found that star formation tends to be more enhanced between closer pairs and pairs with an even mass ratio (Nikolic et al. 2004; Woods \& Geller 2007; Ellison et al. 2008).  Therefore, it is useful to keep in mind the mass ratios and distances between our pair galaxies (our pair galaxies are all galaxies that are gravitationally bound to at least one other galaxy).  In Figure \ref{fig-massvsdist} we plot the mass ratio of the pair against the distance between the two pair galaxies.  The top panel are the V box bound galaxies, and the bottom panel are the C box bond galaxies.  There are some widely separated bound pairs (most of the widely separated pairs are group members), which will result in less tidally-induced star formation.  We will account for this using our distance cuts described above.  There are also some pairs with high mass ratios in both the C and V boxes.  Considering only close bound galaxies reduces the fraction of high mass-ratio pairs, and we find that limiting pairs to more even mass ratios ($\le$ 5) does not qualitatively change any of our results from simply using close bound pairs.

\subsection{Where are Pairs?}\label{sec:wherepair}

As Table \ref{tbl-pairs} quantifies, pairs compose slightly more than half of the galaxy population in the C box and slightly less than half of the galaxy population in the V box (as defined in Section \ref{sec:bounddef}, pairs are all galaxies that are gravitationally bound to at least one other galaxy, so a pair galaxy may be a member of a group).  At all redshift outputs, there is a higher fraction of bound galaxies in the C box than in the V box.  This may have to do with the fact that in the C box there are galaxies that are bound to many other galaxies in groups, while at any output in the V box there are less than half as many groups with at least 4 members as in the C box.  This agrees with the Barton et al. (2007) conclusion that pairs tend to reside in larger dark matter halos (N$_{galaxies}$ $>$ 2).

\begin{figure}
\includegraphics{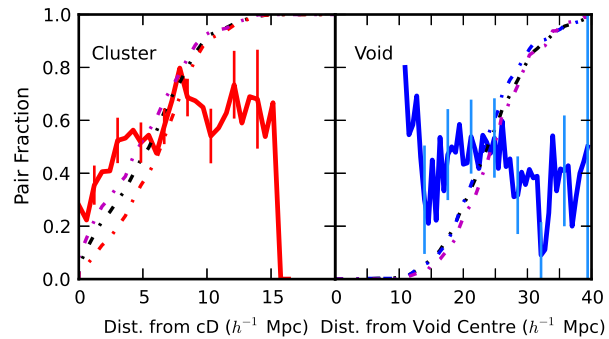}
\caption{This figure illustrates where pairs are in the C and V boxes with respect to the cluster or void centre.  The thick lines are the pair fraction in bins that are 0.6 $h^{-1}$ Mpc wide.  We overplot a few representative binomial error bars (to 95\% certainty).  The dash-dot lines denote the cumulative fraction of different subsets of galaxies.  The black line is the cumulative distribution function of all of the galaxies above the minimum mass in each box.  The red (blue) line denotes the cumulative distribution function (CDF) of the C (V) pairs.  The magneta line denotes the CDF of projected pairs.  Pairs do not dominate the highest or lowest density regions, instead congregating in the groups in the C box.  See Section \ref{sec:wherepair} for discussion.}\label{fig-pairfrac}
\end{figure}

We bin galaxies using either distance from the cluster cD (C box) or void centre (V box), shown in Figure \ref{fig-pairfrac}.  We use the void centre identified in Kreckel et al. (2011).  In this figure we use bin sizes of 0.6 $h^{-1}$ Mpc, and using smaller or larger bin sizes increases or decreases the scatter, respectively, but has no effect on the visible trends.  The solid lines are the pair fraction in different distance bins.  The black dot-dashed line is the cumulative fraction of all galaxies (the total galaxy samples in the C or V boxes) as a function of distance from the cD or void centre.  The red or blue dot-dashed lines are the cumulative fraction of bound galaxies as a function of distance from the cD or void centre.  We also plot the cumulative fraction of galaxies that are members of a projected pair ($\Delta$v $<$ 350 km s$^{-1}$ and r$_{p} <$ 100 kpc $h^{-1}$) as magenta dash-dot lines.  

First we focus on the C box galaxies (left panel).  The total population (black dot-dash line) is concentrated in the cluster--more than 25\% of the galaxies are within 2 Abell radii of the cD (r$_A$ = 1.5 $h^{-1}$ Mpc), and nearly all of the galaxies are less than 10 $h^{-1}$ Mpc from the cD.  

The pair distribution is different from the total galaxy distribution, with few bound galaxies near the cD (the red dash-dot line).  This is not surprising, as galaxies in clusters tend to have large relative velocities.  However, the pair fraction (the thick solid red line) remains nearly constant below 60\% between 3-6 $h^{-1}$ Mpc from the cD, so we do not see evidence of an increase in the fraction of bound galaxies in the infalling population--galaxies are not ``pre-processed" in pairs or groups immediately (within 5 $h^{-1}$ Mpc of the cD) before entering the cluster.  We predict that about 40-50\% of galaxies will enter the cluster in groups or pairs.  At 2 Abell radii the pair fraction is $\sim$ 55\% and has been flat for a few Mpc.  The pair fraction continues to decrease to $\sim$ 37\% at the virial radius.  This range overlaps the Moss (2006) observational finding that 50\%-70\% of galaxies entering clusters are members of pairs \textit{or} have recently merged.  We agree with the semi-analytic results of De Lucia et al. (2011) that 40-60\% of galaxies are isolated when they enter a cluster (also McGee et al. 2009).  We find fewer isolated galaxies entering our cluster than the 70\% in the simulations of Berrier et al. (2009).  Both De Lucia et al. (2011) and McGee et al. (2009) discuss possible reasons for these discrepancies.  Also, our simulations differ in several respects from all three of these works:  we include gas and all hydrodynamical effects as well as radiative cooling and feedback processes, identify galaxies using stellar particles instead of dark matter particles, and define pair galaxies as those that are gravitationally bound.  At larger distances from the cD, 6-15 $h^{-1}$ Mpc, bound galaxies are the majority of the total population.  This is because there are many groups in the C box, in which several galaxies are bound to the central galaxy.  We do not see a similar phenomenon in the cluster because we do not include galaxies bound to the cD in our pair catalogue.  Therefore, 3-6 $h^{-1}$ Mpc from the cD is the closest to an average ``field" population, because it is outside the cluster and groups in the C box.  

Finally, we see that projected pairs (magenta line) are the most centrally-concentrated population of all three.  In dense regions like clusters there is more chance for galaxies to be superposed on the sky but not necessarily bound, which is the case for most projected pairs in our C box (Section \ref{sec:projpairs}).

There are no V box galaxies within 5 $h^{-1}$ Mpc of the void centre, and the first pair is 5 $h^{-1}$ Mpc beyond the first galaxy.  The projected pairs begin farther yet from the void centre.  At large distances from the void centre, one is effectively looking at ``field galaxies".  The galaxy pair fraction in the V box beyond r $\sim$ 15 $h^{-1}$ Mpc is comparable to that of the C box pair fraction between r $\sim$ 3-6 $h^{-1}$ Mpc, which highlights the fact that there is no increased pair fraction near the cluster edge.  The pair fraction in the V box has a large scatter because of the smaller number of galaxies in that region.  Overall, the similarity in the pair fractions in the V box and the "field" region of the C box agrees with the observational result of Szomoru et al. (1996) that small-scale clustering is the same in voids and large-scale higher-density regions.

\begin{figure}
\includegraphics{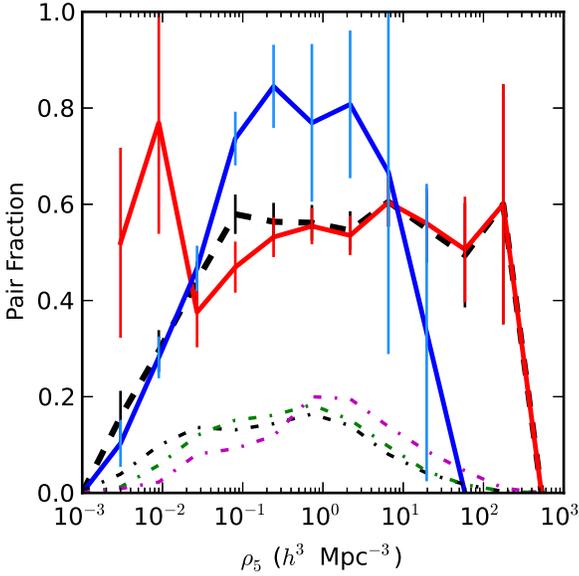}

\caption{The galaxy pair fraction as a function of the local galaxy density.  The thick lines are the pair fraction in local density bins.  We overplot a few representative binomial error bars (to 95\% certainty).As in Figure \ref{fig-pairfrac}, red is C box, blue is V box, and here the black dashed line is the total fraction.  The red spike at low densities is due to the small number of galaxies at low local density in the C box.  The dash-dot lines denote the distribution functions of different subsets of galaxies with respect to local density:  black for the distribution function of the total galaxy population, green for all bound pairs, and magneta for projected pairs.  This figure illustrates that pairs tend to reside in mid-local density regions, particularly avoiding the lowest local densities.  See Section \ref{sec:wherepair} for discussion.}
\label{fig-pairfracld}
\end{figure}

In Figure \ref{fig-pairfracld} we show the pair fraction as a function of local density,
which is defined as $\rho_5 = 5/(4 \pi d_5^3 /3)$, where $d_5$ is the (physical) distance to the fifth nearest neighbor in $h^{-1}$ Mpc.  Notice that this is similar to calculations of local density used by observers (e.g. Perez et al. 2009), but here we calculate a three-dimensional galaxy density. 
As always, we only use galaxies above our minimum stellar mass in the calculation of local galaxy density.  In red we plot the pair fraction in the C box, in blue the V box, and the black dashed line is the total pair fraction across both boxes.  We use large (0.45 dex) bins of local density in order to minimize scatter, but we can still see the effects of small numbers of galaxies in each bin.  
The spike in the pair fraction in the C box at the second-lowest local density bin shown 
is artificially high due to the small number of galaxies in that bin.  Between local densities of 0.1 and 100 galaxies $h^{3}$ Mpc$^{-3}$ the pair fraction in the C box is between 50\% and 60\%, which is somewhat below the pair fraction in the V box between a galaxy density of 0.1 and 3.  This may be because the V box galaxies tend to have more even mass ratios (few extremely massive galaxies at the centres of groups), which will maximize the potential energy between any pair.   

In addition, we plot the distribution of the galaxies across both boxes in each local density bin.  In black we plot the fraction of all galaxies (the total galaxy samples in the C plus V boxes), in green the fraction of all gravitationally bound galaxies, and in magenta the fraction of all projected pair galaxies.  
Several noticeable features are seen.
First, as expected, we see that the total population has a broader distribution of local densities than either the bound or projected pair galaxies, and a much larger fraction of the total galaxy population is at lower local densities ($\rho_5$ $<$ 0.1 $h^{3}$ Mpc$^{-3}$) than that at which bound or projected pairs are found.  Bound galaxies (green) are rarely found in the lowest density regions, and drop at least as sharply as the total population at the highest densities ($\rho_5$ $>$ 10 $h^{3}$ Mpc$^{-3}$).  Projected pairs (magenta) are also rare at low densities.  However, the projected pair distribution peaks at densities that are higher than either the total or bound population, and a higher fraction of the projected pair population is found at the highest local densities.

\subsection{Wet, Dry, and Mixed Mergers}\label{sec:wdm}

We now ask where wet, dry and mixed mergers occur.  Given the colour-magnitude diagram in Figure \ref{fig-colourhist}, it is sensible to divide the whole population into red and blue sequences at $g-r$ = 0.65, as in observational work (e.g. Lin et al. 2008; Patton et al. 2011 use a line with slope -0.01 that passed through $g-r$ = 0.65 at M$_r$ = -21).  With this colour cut we denote blue-blue pairs as wet, red-red as dry, and blue-red as mixed (pairs are defined in Section \ref{sec:bounddef} to be galaxies gravitationally bound to at least one other galaxy).  We find that 59\% of pairs are wet, 13\% are dry, and 28\% are mixed (C box: 48\% wet pairs, 17\% dry pairs, 35\% mixed pairs;  V box: 91\% wet pairs, $<$1\% dry pairs, 9\% mixed pairs -- the numbers separated by redshift and box are in Table \ref{tbl-wdm}).  The ratios between the types of pairs are in good agreement with the observational results of Lin et al. (2010), who find 56\% wet pairs, 15\% dry pairs, and 29\% mixed pairs.  However, there are differences between our sample and the Lin et al. (2010) sample.  For example, we have a broader range of local density.  If we use the same range of local density relative to our median $\rho_5$ (0.45) as in Lin et al. (2010) (one order of magnitude below and two orders of magnitude above the median), we find 60\% wet pairs, 11\% dry pairs, and 29\% mixed pairs, still consistent with the Lin et al. (2010) fractions.   The Lin et al. (2010) sample includes galaxies with -21 $<$ M$_B$+1.3 $<$ -19, whereas we include massive galaxies above their range.  As we have discussed in Section 2.2, star formation is over-estimated for the most massive galaxies, and our slightly higher fraction of wet pairs may be due to this.

\begin{table}
\begin{center}
\caption{Number of Wet, Dry, and Mixed Pairs\label{tbl-wdm}}
\begin{tabular}{c | c | c | c | c}
\hline
Box & redshift & wet & dry & mixed\\
\hline
C &  0 & 88 & 42 & 56 \\
C & 0.05 & 154 & 45 & 126\\
C & 0.1 & 173 & 84 & 138\\
C & 0.15 & 189 & 61 & 127\\
C & 0.2 & 212 & 54 & 136\\
V & 0 & 42 & 0 & 5\\
V & 0.05 & 133 & 0 & 14\\
V & 0.15 & 154 & 1 & 18\\
V & 0.2 & 171 & 0 & 11\\
\hline
\end{tabular}
\end{center}
\end{table}

\begin{figure}
\includegraphics{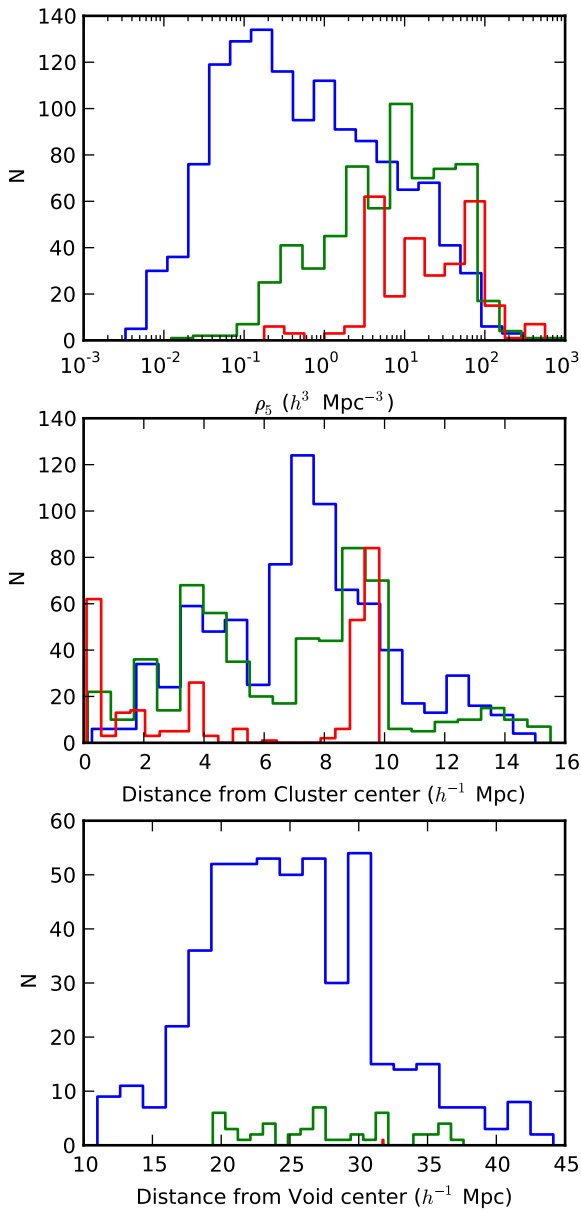}
\caption{Histograms of pairs as a function of local density and distance from the cD or the void centre.  Red denotes dry pairs, blue denotes wet pairs, and green denotes mixed pairs.  The differences in the distributions of dry and wet pairs are easy to pick out by eye.  See Section \ref{sec:wdm} for discussion.}
\label{fig-wdmhists}
\end{figure}

In Figure \ref{fig-wdmhists} we plot the histograms of the wet (blue), mixed (green), and dry (red) pairs as a function of distance from the cluster centre or void centre, and of local galaxy density.   Both the clustercentric distance and local density of the major member of the pair are used to represent those of the pair.  For this reason it is easy to pick out the two largest groups in the C box in the middle panel--at about 8 $h^{-1}$ Mpc and 9-9.8 $h^{-1}$ Mpc from the cluster centre.  One of them has a red central galaxy and therefore the pairs are either dry or mixed (at $\sim$ 9.8 $h^{-1}$ Mpc), and the other has a blue central galaxy so the pairs are either wet or mixed (at $\sim$ 8 $h^{-1}$ Mpc).  The group with the blue central galaxy has more blue galaxies, and the group with the red central galaxy has more red galaxies (seen in the top two panels of Figure \ref{fig-wdmhists}).  This colour conformity is consistent with the 'galactic conformity' observed in Weinmann et al. (2006), that the morphology of the central galaxy in a halo is correlated with the morphology of the satellite galaxies (this was recently examined in terms of gas availability in Kauffmann et al. 2010).

Ignoring these two groups, we can see that dry pairs are closer to the cluster centre, in general agreement with the colour-density relation.  Wet and mixed pairs are spread much more evenly in distance from the cD, although there are very few wet pairs within the virial radius of the cluster (1.3 $h^{-1}$ Mpc), and no wet pairs within the virial radius at z $\le$ 0.05.  In the V box (the lowest panel), there are no pairs within 10 $h^{-1}$ Mpc of the void centre, and no mixed pairs until nearly 20 $h^{-1}$ Mpc from the void centre.  The number of pairs increases with distance from the void centre until we reach the edge of the V box.

If we consider the local galaxy density of the pairs in the upper panel of Figure \ref{fig-wdmhists}, we can more easily pick out the group with the red central galaxy, which has a local density of about 10 $h^3$ Mpc$^{-3}$.  At later times (at z$\le$0.05), both groups have this local density.  However, at earlier times the red group has even higher local densities and the blue group has lower local densities.  Focusing on the rest of the pairs, we see that wet pairs reach the lowest local densities ($\rho_5 \sim$ 0.005).  We also find that dry pairs are rare outside of group environments ($\rho_5 <$ 3), although we might find more dry pairs if we had a larger red galaxy population (if, for example, we included AGN feedback, see Section \ref{sec:galaxyselection}).  Our overall finding is that, while all three types of pairs can occur in groups, dry pairs are extremely rare in field environments.  If pairs end in mergers, the following physical picture emerges:  dry mergers should occur only in group or cluster environments, while wet and mixed mergers can occur at all local galaxy densities.  Wet mergers will dominate at the lowest local densities ($\rho_5 <$ 0.1).  At the other environmental extreme, dry mergers dominate within the virial radius of a cluster.  This is in general agreement with the observations of Lin et al. (2010).

\section{Pairs Compared to the Total Galaxy Population}\label{sec:comparisons}

In this section we compare properties of pair galaxies (we define pairs in Section \ref{sec:bounddef} to be galaxies bound to at least one other galaxy--excluding the cD in our largest cluster) to those of the entire galaxy population.  First consider the stellar mass of the galaxies.  We find that, on average, the stellar mass of bound galaxies is larger than that of the total population at a fixed local density.  This is true whatever subset of bound galaxies we consider (all pairs, d $<$ 500 $h^{-1}$ kpc, d $<$ 250 $h^{-1}$ kpc, or the HOP multi-peak sample).  Because galaxy mass may have a strong influence on galaxy properties (Visvanathan \& Sandage 1977; Schweizer \& Seitzer 1992; Blanton et al. 2003; Baldry et al. 2004; and see Figure \ref{fig-colourhist}), we performed all of our analysis twice:  once with the entire galaxy sample as our control, and once using a sample that was matched to the stellar mass distribution of the bound sample.  Each bound sample was separately matched, therefore the comparison samples are not identical.  After going through this process we find that the differences between the pair and total samples are somewhat smaller when we match stellar mass, although the sign remains the same.  In this section, we will present our results using the entire galaxy sample for comparison.

\subsection{Star Formation Rate}\label{sec:colour}

Galaxy colour is often used as a proxy for star formation rate (see the references in the Introduction).  We have considered both the colour and specific SFR (sSFR $\equiv$ SFR/M$_*$), and find analogous results when comparing pairs to the total population.  Therefore we will focus on the sSFR, as it is a more fundamental parameter and calculated directly in the simulation, whereas colour is subject to other less well modeled processes, such as dust reddening.  We will now discuss the differences in the sSFR between the pair and the total galaxy population.  As we discussed in Section \ref{sec:galaxyselection}, we will compare trends we find in our simulations to observational trends rather than attempt to directly compare our simulated galaxies to observed galaxy populations.   

\begin{figure*}
\includegraphics{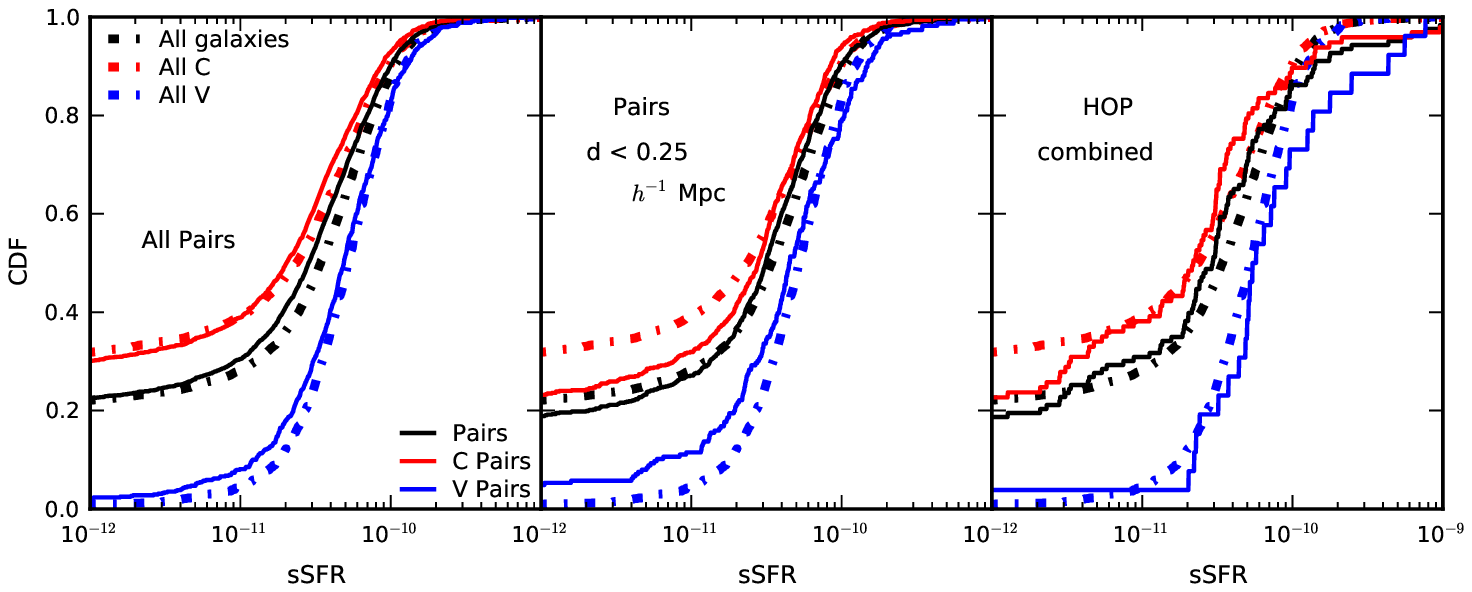}\\
\includegraphics{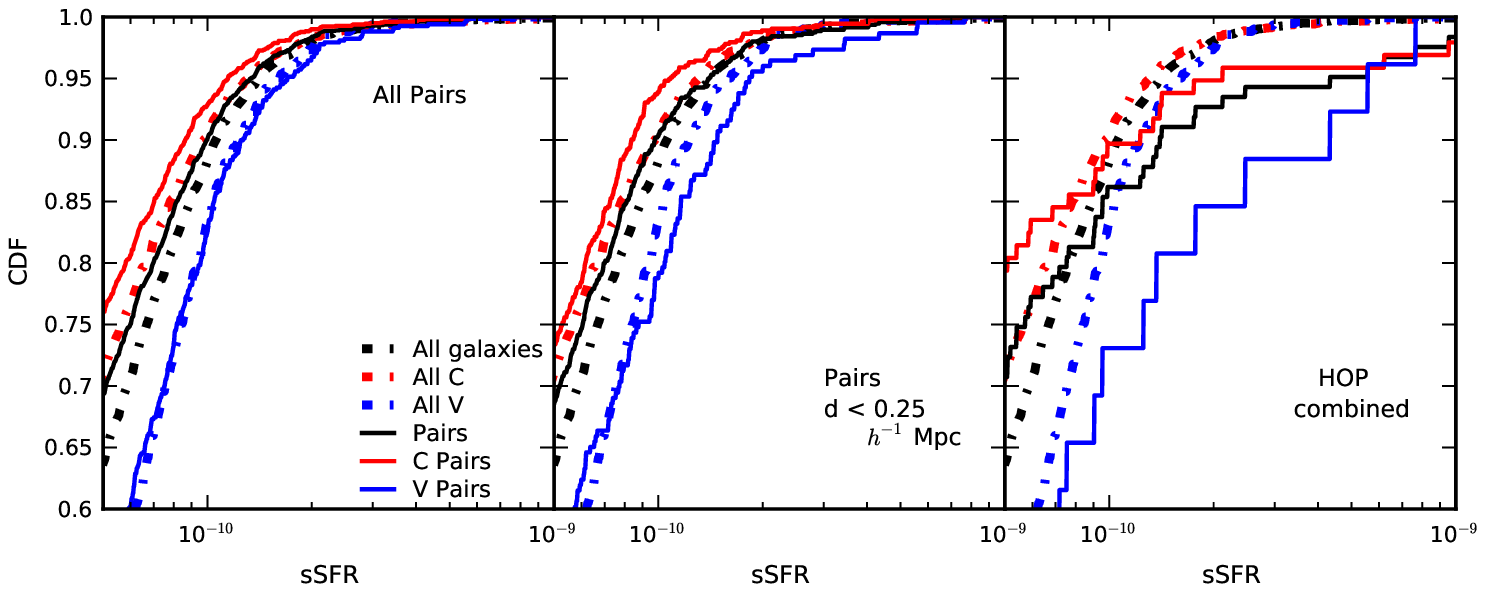}
\caption{Top three panels show the sSFR distribution of the galaxies.  The CDFs for the total galaxy population in the C box, V box, and across both the C and V boxes are shown in all three panels as the red, blue, and black dash-dotted lines, respectively.  Left panel shows the CDF for all bound galaxy pairs.  The middle panel shows a subset of those pairs are that less than 250 $h^{-1}$ kpc apart.  The right panel shows galaxies that were identified by HOP as a single galaxy because their stellar clumps significantly overlap.  The bottom three panels zoom in to the high sSFR end.  Note that the galaxies in the C box have a different sSFR CDF than galaxies in the V box (Section \ref{sec:colour}).  Closer bound galaxies have a higher sSFR than more distant ones, and close pairs (d $<$ 250 $h^{-1}$ kpc) have a slightly higher fraction of galaxies with sSFR $>$ 10$^{-10}$ yr$^{-1}$ than galaxies in the total population.  HOP multi-peak galaxies have a higher fraction of galaxies with high sSFRs, and the C HOP multi-peak galaxies also have fewer galaxies with low sSFRs (See Section \ref{sec:colour}).}
\label{fig-ssfr}
\end{figure*}

In Figure \ref{fig-ssfr} we plot the cumulative distribution function (CDF) of the sSFR of different subsets of the simulated galaxies.  The dash-dotted lines remain the same in all six panels, describing the total galaxy population in the C box (red), V box (blue) and including the total combined population in the C and V boxes (black).  Overlaid on these curves we plot the CDFs of bound galaxies, with each panel showing a different subset.  To make the distributions of the higher sSFR galaxies more clear, we zoom in on the high sSFR end in the bottom panels.

The sSFR distributions of the total galaxy populations in the two boxes differ.  The V galaxies have a much higher fraction of high sSFR galaxies, although the maximum sSFR in both boxes is very similar.  About 35\% more of the galaxies in the V box have sSFR $>$ 10$^{-11}$ yr$^{-1}$ than in the C box.  Clearly sSFR (and therefore galaxy colour) depends on environment, a point we will return to in Section \ref{sec:env}.

Note that the bound (and to a lesser extent, the total galaxy) CDFs are dominated by the C box galaxies, given the numbers shown in Table \ref{tbl-pairs}.  Focusing on the bound galaxies, we see that the CDF tilts towards higher sSFR as we move from the entire bound sample to subsets with distance cuts.  This is largely due to changes in the CDF of the C box in the low sSFR end ($<$ 5 $\times$ 10$^{-11}$).  In the lower panels, it is clear that there is very little change in the distribution of higher sSFR galaxies between the C box total bound population (0.9$|$W$|$ $>$ K) and the C box bound population with d $<$ 250 $h^{-1}$ kpc.  

If we focus on the V box, we find that the bound galaxy distribution appears very similar to that of the total V galaxy population, but with slightly more galaxies at both the high and low sSFR ends.  In contrast to the C box galaxies discussed above, the sSFR distribution of the low sSFR V box bound galaxies varies very little as we consider only pairs with small separations ($\sim$2\% at 10$^{-11}$ yr$^{-1}$).  The most noticeable change in the V pair CDFs is a larger fraction of high sSFR galaxies (+5\% at 10$^{-10}$ yr$^{-1}$). 

The HOP multi-peak sample has a clearly larger high sSFR fraction than the total population for both the C and V boxes, although it is more pronounced in the V box.  The sSFRs in these strongly interacting galaxies are among the highest of the total galaxy population.  There are also galaxies with low sSFRs in the HOP multi-peak sample, indicating that being a member of a strongly interacting pair does not necessarily lead to strong star formation.  When we match this sample for stellar mass we see the exact same trends.  

Tidal interactions may be causing a substantially higher fraction of HOP multi-peak galaxies to have high sSFRs than pairs with larger separations.  Of course, companion-induced interactions are not the only cause of blue galaxies with high sSFRs; in fact, the galaxy with the highest sSFR (1.75 $\times$ 10$^{-9}$ yr$^{-1}$) is not bound to any other galaxy.  Quantitatively, among the 39 galaxies with the highest sSFRs ($\ge$ 3 $\times$ 10$^{10}$ yr$^{-1}$), we find only 8 of them to be members of the bound d $<$ 250 $h^{-1}$ kpc, 6 to be a member of the HOP multi-peak sample, and 1 to have been a member of the HOP multi-peak sample in the previous output (and therefore likely to have just merged).  In Figure \ref{fig-ssfr} we find red HOP multi-peak galaxies with low sSFRs that span the entire stellar mass range and the entire range of mass ratios in the HOP multi-peak sample.  

To briefly summarize, we find a slightly higher fraction of low sSFR bound galaxies than that of the total sample (compare the black dash-dotted line to the black solid line in the first panel of Figure \ref{fig-ssfr}).  This is in agreement with Patton et al. (2011).  Including any distance cut results in a pair population with higher sSFR.  This is at least partly because the distance cuts remove more group galaxies, which are likely to have lower sSFR (and therefore be redder) due to the dense environment in which they reside.  We find a larger population of high sSFR (extremely blue) galaxies in the HOP multi-peak pairs than in the total sample, which qualitatively agrees with the results of Patton et al. (2011).  We do not find a larger fraction of low sSFR (extremely red) galaxies using any of our pair samples.  

\subsection{Cold Gas Mass}\label{sec:sfrgas}

\begin{figure}
\includegraphics{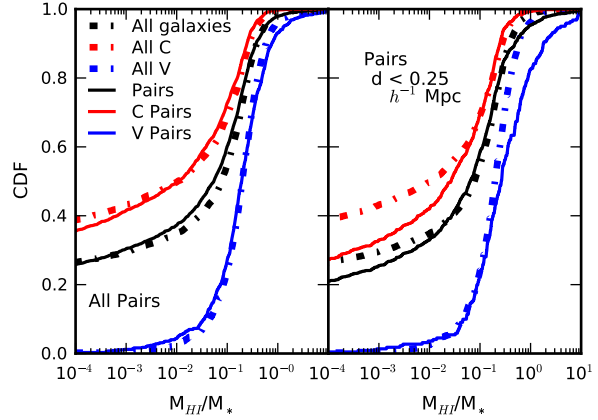}
\caption{Left panel shows the M$_{HI}$/M$_*$ distribution of the galaxies, for pairs and the total galaxy population in the C box, V box, and across both the C and V boxes.  Right panel shows the CDFs for closer pairs with d $<$ 250 $h^{-1}$ kpc.  Linestyles are as in Figure \ref{fig-ssfr}.  See Section \ref{sec:sfrgas} for discussion.}  
\label{fig-smhi}
\end{figure}

In addition to determining whether star formation is affected by being a member of a pair or group, 
we now investigate whether the cold gas mass of bound galaxies is different from the total population.  As mentioned in Section 2.1, the \ion{H}{I} density is directly followed within the simulation.  We consider the \ion{H}{I} gas fraction in Figure \ref{fig-smhi}, as it tells us about the fuel available for star formation in these galaxies.  We use the \ion{H}{I} gas mass within one virial radius.  We are unable to accurately separate the gas that belonged to each component galaxy in the HOP multi-peak sample, and so do not consider that sample in this section.

In the C box, the \ion{H}{I} gas fraction for all pairs has a similar distribution to the general C population.  It is notable that close pairs have a higher fraction (70\% vs 60\%) of somewhat \ion{H}{I} gas rich galaxies (M$_{HI}$/M$_*$ $\ge$ 10$^{-3}$) than the total C population.  It is equally interesting to note that the fraction of pairs ($\sim$4\%) with high \ion{H}{I} mass fractions (M$_{HI}$/M$_*$ $\ge$ 0.1) appears little changed by being a pair member or a close pair member.  

As in the C box, V box pairs have a very similar \ion{H}{I} gas fraction to the general V population.  However, unlike the C box pairs, bound pairs in the V box with smaller separations (d $<$ 250 $h^{-1}$ kpc) have a dramatically higher fraction of extremely \ion{H}{I} rich galaxies:  20\% of close pairs have M$_{HI}$/M$_*$ $\ge$ 1, whereas only 4-5\% of galaxies in the total V population or the entire V pair population have M$_{HI}$/M$_*$ $\ge$ 1.  

Overall, we find that galaxies in closer pairs have higher M$_{HI}$ gas fractions than the general population:  the P values for the KS tests comparing the bound galaxies to either the total or the mass-matched sample are below 0.1 for all pairs with d $<$ 500 $h^{-1}$ kpc.  This indicates that close interactions increase cold gas formation via gravitationally induced hydrodynamical effects and radiative cooling.  Observations indicate that galaxy interactions can cause inflows in ionized gas (Rampazzo et al. 2005) and neutral gas (e.g. Hibbard \& van Gorkom 1996).  In addition, observations find that interacting galaxies have lower metallicities in the central regions and have lower metallicity gradients in the disc (Kewley et al. 2006; Kewley et al. 2010).  Together these findings indicate that low metallicty gas inflows come from outer regions of the halo to the disk and central galaxy.

Recall from Figure \ref{fig-ssfr} that the fraction of higher-sSFR galaxies does not start to increase until we focus on the HOP multi-peak sample.  This may mean that gas is cooling from the halo in the d $<$ 250 $h^{-1}$ kpc sample, but is not transmitted into a higher SFR until galaxies are still closer.  Our simulations suggest this physical picture holds for about 50\% of the close pairs (d $<$ 250 $h^{-1}$ kpc) in the V box with M$_{HI}$/M$_*$ $\ge$ 0.2.

\section{The Local Environment of Bound Galaxies}\label{sec:env}

In higher density environments, observed galaxies tend to be more massive, redder (lower sSFR), and have less \ion{H}{I} gas (Hubble \& Humason 1931; Oemler 1974; Dressler 1980; Balogh 2001; Blanton et al. 2003; Solanes et al. 2001; Haynes, Giovanelli, \& Chincarini 1984).  By simply comparing the total galaxy population in the C box to that in the V box, it is clear that our simulated galaxies follow these relationships well (e.g. Figures \ref{fig-ssfr} and \ref{fig-smhi}).   

\begin{figure*}
\includegraphics{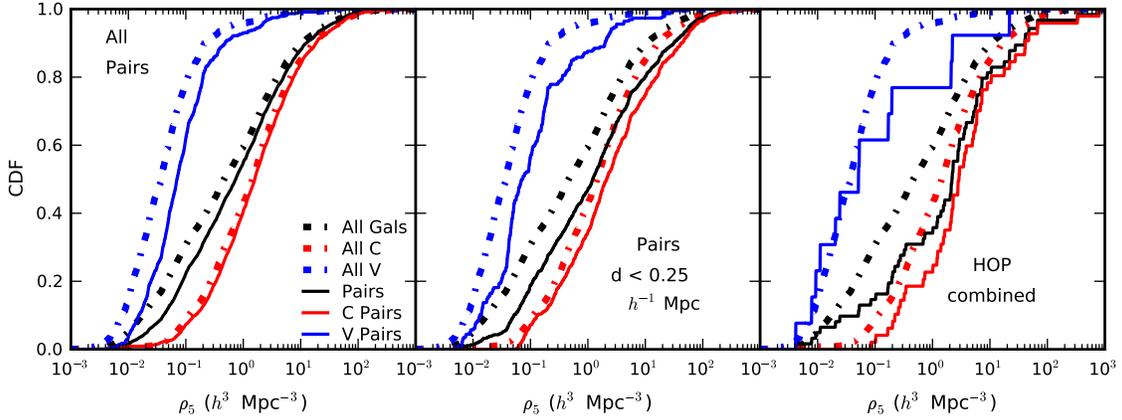}
\caption{CDFs of the local density, $\rho_5$, of galaxies.  See Figure \ref{fig-ssfr} for a description of the panels.  Bound galaxies tend to reside in local densities above that of the corresponding total population.  See Section \ref{sec:env} for discussion.}\label{fig-localdens}
\end{figure*}

Therefore, we must determine if the differences between pairs (as defined in Section \ref{sec:bounddef}, a pair galaxy is gravitationally bound to at least one other galaxy that is not the cD of the largest cluster) and the total galaxy population may be attributed to an environmental difference between bound and unbound galaxies.  We show in Figure \ref{fig-pairfrac} that there are fewer pairs in regions close to the cD ($<$ 3 $h^{-1}$ Mpc) than in regions more distant from the cluster.  This may be part of the reason that the bound galaxies in this large scale environment have fewer galaxies with low sSFRs than the general galaxy population in the C box:  simply because pairs are more remote from the cluster (see Figure \ref{fig-ssfr}).  Similarly, we find that bound galaxies in the V box tend to be farther from the void centre than the general population.  This could explain why bound galaxies in this low-density large-scale environment tend to have lower sSFRs than the general V galaxy population--although very close bound galaxies in the V box do not conform to this trend (Figure \ref{fig-ssfr}).  Even though pairs avoid the centre of the cluster and void, Figure \ref{fig-pairfracld} shows that in both the C and V boxes the local galaxy density of bound galaxies spans a large range.  

To look more closely at the local environment of bound galaxies versus the total population, Figure \ref{fig-localdens} shows the CDFs of our galaxy pair and total populations.  We see that V box pairs tend to reside in regions of higher density (a factor of $\sim$2) than the total V population, whereas in the C box the environments are nearly identical, bearing mind that bound pairs with the cD galaxy of the cluster are intentionally removed from the bound pair populations in the C box in our analysis..  The closest pairs (d $<$ 250 kpc $h^{-1}$) in both the C and V boxes reside in a higher range of local galaxy density than the total samples.  This is also true if we compare the bound galaxies to the total sample that is matched in mass.  The only KS test with a P value larger than 0.1 compares the total C population matched in mass to all the pairs in the C box.  Clearly, bound galaxies tend to reside in higher local density environments than unbound galaxies.  The fact that there is a higher fraction of close pairs than pairs with larger separations in high-density environments suggests either that dense environments are more conducive to the formation of close pairs, or that in general pairs have a lifetime on order of the Hubble time and move closer together as they migrate to denser environments, or a combination of these two factors.

In the C box, bound galaxies have a strong tendency to have higher sSFRs and be more \ion{H}{I}-rich, which is opposite to what one would predict from the fact that bound galaxies reside in higher density environments.  It is less clear whether the V box bound galaxies are influenced by the fact that they tend to reside in higher density environments than the total population.  First, there is an excess of galaxies with low sSFRs in all V bound galaxy populations except the HOP multi-peak sample.  This trend could be explained by the environment.  However, in the d $<$ 250 kpc $h^{-1}$ galaxies and the HOP multi-peak galaxies there is an excess of high sSFR galaxies as well.  The V bound population tends to have more \ion{H}{I}-rich galaxies, which would not be caused by a dense environment.  Combining these facts, it seems that most of the differences between the bound and total galaxy populations may be driven by pair effects rather than local environment.  We will now test this more carefully.

\begin{figure}
\includegraphics{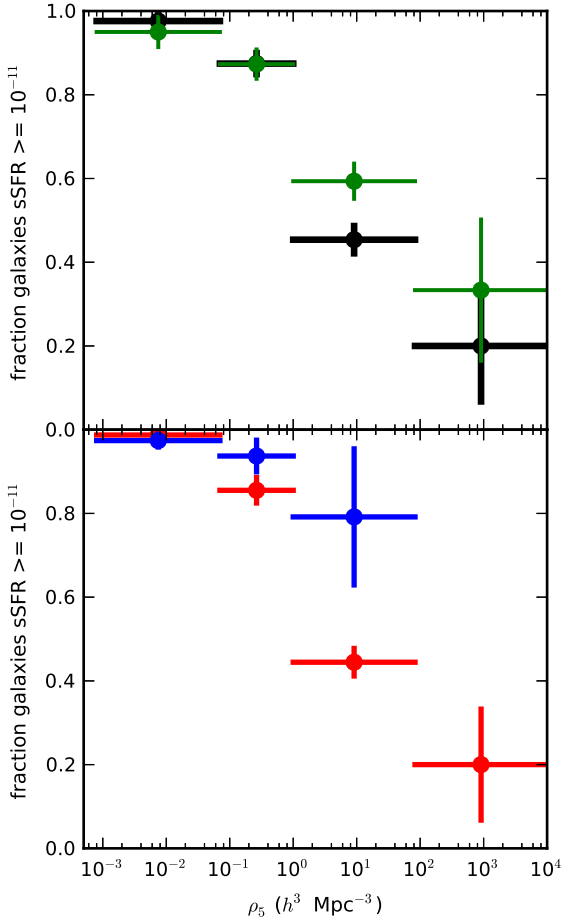}
\caption{The fraction of star-forming galaxies in four bins of local density.  \textbf{Top Panel:}  the black points are the star-forming fractions of the total comparison population, and the green lines are the star-forming fraction of the close pairs (d $<$ 250 kpc $h^{-1}$ galaxies plus HOP multi-peak galaxies) for the combined C plus V boxes.  \textbf{Bottom Panel:}  The red points are the fractions of mass-matched C box total galaxy sample, and the blue points are those of the mass-matched total V box sample.  The horizontal lines denote the width of the local density bin, and the vertical lines are the 95\% confidence binomial error bars.   Note the only clear difference between pairs and the matched sample is in the star-forming fraction at the highest densities.  See Section \ref{sec:env} for discussion. }\label{fig-bluefrac}  
\end{figure}

\begin{figure}
\includegraphics{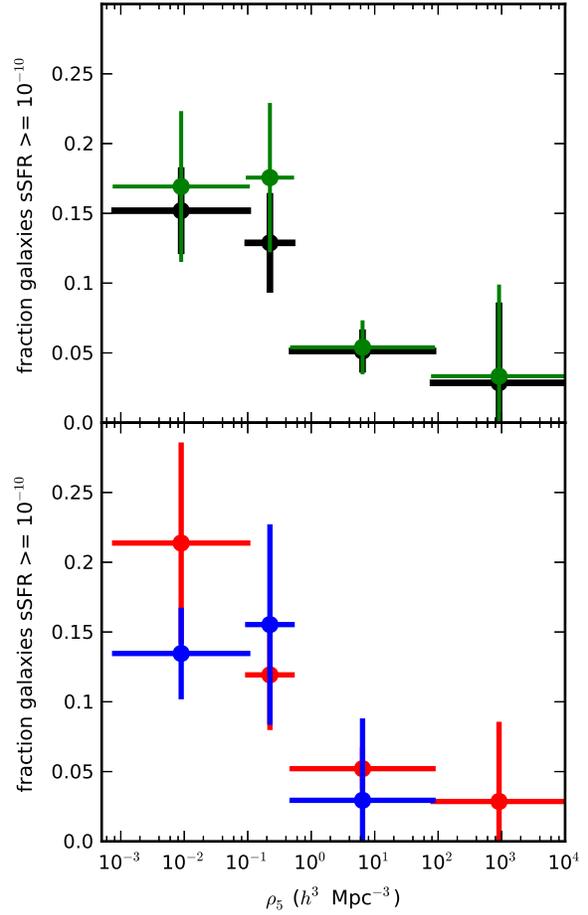}
\caption{The fraction of starburst galaxies in close pairs (d $<$ 250 kpc $h^{-1}$ galaxies plus HOP multi-peak galaxies) compared to a sample of the total population matched to the mass distribution of the pairs.  See Figure \ref{fig-bluefrac} for an explanation of the symbols.  See Section \ref{sec:env} for discussion. }\label{fig-verybluefrac}  
\end{figure}

\subsection{Fraction of Star-Forming Galaxies Relative to Their Environment}

We can look more closely at the effects of the large-scale, local, or very local (being a member of a close bound pair, where a pair galaxy is any galaxy that is gravitationally bound to at least one other galaxy that is not the cD of the largest cluster) environment on the star-formation properties of galaxies.  We define the fraction of galaxies having sSFR $>$ 10$^{-11}$ yr$^{-1}$ as the star-forming fraction.  This is the lowest sSFR denoting the green valley used by Heinis et al. (2009).  In the upper panel of Figure \ref{fig-bluefrac} we plot the star-forming fraction in four bins of local density, which were chosen so that the second bin contained a large number of galaxies from both the C and V boxes, and the third bin extends to the highest local density in the V box ($\sim$83 $h^3$ Mpc$^{-3}$).  The comparison ``total" population has been matched to the mass distribution of the close pair sample.  We use close pairs (d $<$ 250 kpc $h^{-1}$ including the HOP multi-peak sample).  The horizontal lines are the widths of the local density bins, and vertical lines denote the binomial error for the sample in that bin using a 95\% confidence level.  

In the lower panel of Figure \ref{fig-bluefrac}, we plot the total galaxy population from each box separately.  The C box total population are the red symbols, and the V box total population are the blue symbols.  We note that we have a small number of galaxies in the lowest and highest density bins from the C box, and in the third density bin from the V box.     

We can make several comparisons using Figure \ref{fig-bluefrac}.  First, we can determine how different local densities affect the star-forming fraction of the total galaxy population by comparing the black symbols.  Second, we can determine how the large-scale structure affects the star-forming fraction by comparing the solid red (C box) and blue (V box) markers.  Finally, we can check how being a member of a pair affects the star-forming fraction by comparing the green and black symbols in the top panel.

Addressing the first point, we find that the fraction of star-forming galaxies decreases with increasing local density.  This is true whether we are considering the total galaxy population (black symbols), or the total pair population (green symbols).  This is also true when considering the C box or V box galaxies separately (bottom panel).  This is in excellent agreement with observations where there is a consistent trend of galaxies becoming progressively redder with lower sSFRs from voids to filaments to clusters (Rojas et al. 2004, 2005; Kauffman et al. 2004; G{\' o}mez et al. 2003).  

Focusing on the C and V box galaxies separately leads us to our second question:  how does the large-scale structure affect the star-forming fraction?  In the bottom panel, the trend with local density is much stronger for the C box galaxies, which drops by about 55\% compared to the $\sim$20\% drop in the V box over the same local density range (the first three bins).  This drop remains nearly identical (30\% difference versus 35\% difference) when we match the local density distributions (P value of 0.93) and the mass distributions (P value of 0.78) of the C and V galaxies within the third density bin (the bin with the largest difference between the C and V box total galaxy populations).  In the two lower density bins, the V box and C box galaxies have similar star-forming fractions, and the V box galaxies are less affected by local density than the C box galaxies only in the third density bin, where the error bars on the V box galaxies are quite large.  The lower star-forming fraction at similar $\rho_5$ (1 $<$ $\rho_5$ $<$ 83) may indicate that C box galaxies are more likely to be in larger groups than V box galaxies.  Although to first order $\rho_5$ is a measure of halo mass, in large halos there can be a large range of $\rho_5$.  For example, even within one Abell radius of the cluster centre (1.5 $h^{-1}$ Mpc), $\rho_5$ can be as low as 1.5 $h^3$ Mpc$^{-3}$.  We find that in general the local density has a stronger affect on the star-forming fraction than whether a galaxy is in the C box (large-scale overdensity) or V box (large-scale underdensity).  However, at higher local densities, there is a significant difference between C and V box galaxies, which may be indication of some effects on intermediate scales (~1 Mpc), such as whether or not a galaxy is close to the cluster (or a group). 

Finally, we consider whether being a member of a pair affects the star-forming fraction.  In the top panel we compare the total population (black) to the close pairs (d $<$ 250 kpc $h^{-1}$ including HOP multi-peak galaxies in green) in the C plus V boxes.  In the two lowest density bins, the populations have the same star-forming fraction within the error bars.  In the two highest local density bins, the star-forming fractions of the pairs are higher than the star-forming fraction of the total population (although the error bars in the highest bin are large).   

We checked these results using only the HOP multi-peak sample, and while the error bars are much larger, we find the same trends.  We also considered whether simply being near another galaxy could cause this increase in the star-forming fraction by comparing the close bound galaxies to the close unbound galaxies (see Appendix \ref{sec:unbound}).  Although with four close unbound galaxies in the V box our results are dominated by the C box galaxies, we still find the same trends--the star-forming fraction decreases with local density and in the two lower-density bins the bound galaxy star-forming fraction agrees within the errors with that of the close unbound galaxies.  In the highest local density bin the bound galaxy star-forming fraction is significantly higher than that of unbound close galaxies.

We have also examined the median sSFR of star-forming galaxies in Appendix \ref{sec:medianssfr}. We find that the star-forming fraction does not strongly depend on galaxy mass and the local density distribution of pairs versus the total sample in Appendices \ref{sec:ssfrmass} \& \ref{sec:ssfrlocald}.

To summarize all of the information in Figure \ref{fig-bluefrac}, we find two main effects on the star-forming fraction.  First, the fraction of star-forming galaxies is most strongly dependent on local density ($\rho_5$), dropping by about 80\% (from nearly 100\% to about 20\%) from our lowest to highest density bin (i.e. from voids to clusters).  Second, being a member of a bound pair affects the star-forming fraction of the galaxy population only at the highest local densities, with a boost in the star-forming fraction of more than 10\%.  

We believe the first conclusion to be very robust and it is indeed a fundamental environmental effect.  The second effect is more difficult to decipher physically for the following reason.  It could be that the local environment indeed plays the primary role to produce the seen effect.  Alternatively, one might envision the following picture.  Galaxies tend to move to higher local density environments with time, whether they are isolated, in pairs, or in groups.  Pairs may form at some intermediate local density.  With time these pairs move to higher density regions and at the same time become more closely separated.  If a portion of these close pairs results in more enhanced star formation, it could explain, at least in part, the seen trend.  Such a picture is also consistent with the higher fraction of close pairs in higher density environments (Figure \ref{fig-localdens}).  However, we would expect the HOP multi-peak sample to have a higher fraction of star-forming galaxies at any local density, and we only find a difference of that sort in the highest local density bin.  We defer a more careful examination of this issue to a future study.

\subsection{Fraction of Starburst Galaxies Relative to Their Environment}

In Figure \ref{fig-verybluefrac} we plot the fraction of galaxies with sSFR $>$ 10$^{-10}$ yr$^{-1}$ in four bins of local density, which as before are chosen so that the second lowest density bin contained a large number of galaxies from both the C and V boxes, and the second highest density bin reached the maximum local density of V box galaxies.  This sSFR is the limit used by Heinis et al. (2009) to denote the bottom of the blue cloud, which we call starburst galaxies.  If we compare close pairs (d $<$ 250 $h^{-1}$ kpc bound including HOP multi-peak galaxies) to the mass-matched total galaxy sample, we can clearly see that the values are the same to well within the errors.  In fact, these errors are large enough that we can only say that the fraction of starburst galaxies decreases with increasing local density (by about 10\%).  The fraction of starburst galaxies is affected less than the fraction of star-forming galaxies by both local density and whether a galaxy is a member of a close pair. 

\begin{figure}
\includegraphics{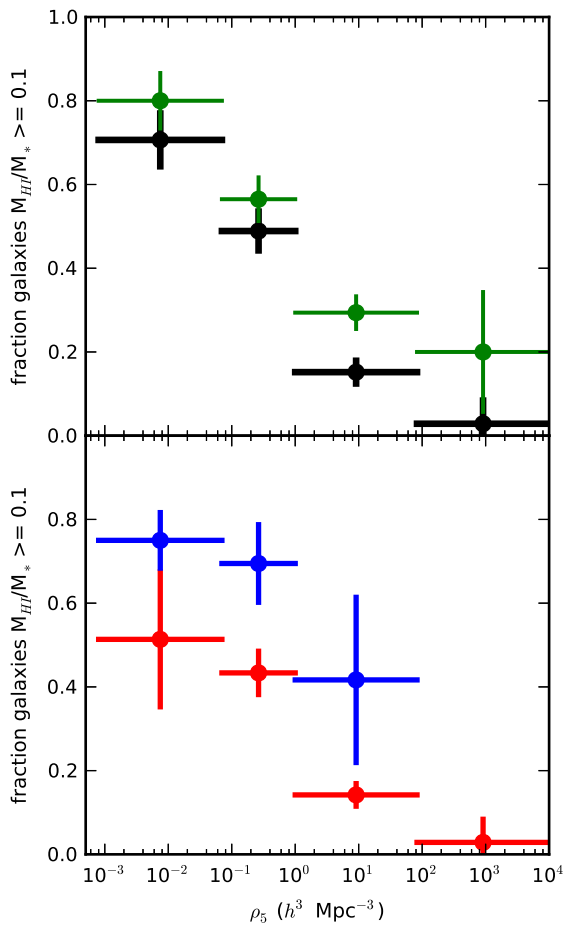}
\caption{The fraction of \ion{H}{I}-rich galaxies in close pairs (d $<$ 250 kpc $h^{-1}$ galaxies plus HOP multi-peak galaxies) compared to a sample of the total population matched to the mass and local density distribution of the pairs.  The symbols are the same as in Figure \ref{fig-bluefrac}.  Once again, the samples differ significantly in the two highest local density bins.  See Section \ref{sec:env} for discussion.}\label{fig-mhifrac}
\end{figure}

\subsection{Fraction of \ion{H}{I}-Rich Galaxies Relative to Their Environment} 

In Figure \ref{fig-mhifrac} we plot the fraction of galaxies with M$_{HI}$/M$_*$ $>=$ 0.1, which we will call the \ion{H}{I}-rich fraction, with the comparison sample matched for both mass and local density to the pair sample.  As with the star-forming fraction, the bound galaxies differ from the comparison population only in the highest local density bins.  The pairs have a higher \ion{H}{I}-rich fraction, which is consistent with having a higher star-forming fraction of galaxies, as seen earlier.

But, unlike the star-forming fraction, the \ion{H}{I}-rich fraction is consistently higher in the V box than in the C box.  This indicates that galaxies in the V box have easier access to cold gas, which, by comparison to the star-forming fraction, does not necessarily result in high star-formation rates.  

\subsection{Comparison to Observations}

We will now briefly compare our results to the observations of Ellison et al. (2010).  Ellison et al. (2010) consider whether environment changes the effect of being in a close pair.  In direct opposition to our results, they find that star formation is more enhanced for close pairs at low projected local density, with no increase in the median sSFR for projected pairs in their highest local density bin.  Using projected local density, we still find results opposite to theirs.  A number of factors may have contributed to this difference.  

The main physical difference between our simulation and their observation is that pairs in our simulation are physically bound, while observed projected pairs may contain interlopers.  In particular, the paired galaxies in the higher density bins in Ellison et al. (2010) may contain a high fraction of unpaired interlopers (as we have shown earlier), which may lower the sSFR of their pair population in high density regions.  To illustrate how this affects our results, we find that the fraction of star-forming projected pairs in our third local density bin (1 $<$ $\rho_5$ $<$ 83) is 49\%, which is below the 60\% star-forming fraction of close pairs (d $<$ 250 kpc $h^{-1}$ galaxies plus HOP multi-peak galaxies), and within the error bar of the total galaxy population shown in the top panel of Figure \ref{fig-bluefrac}.  

The sample used in Ellison et al. (2010) is much larger than our sample, so the close pairs at which they see this increase in sSFR are within 30 kpc $h_{70}^{-1}$.  This corresponds roughly to our HOP multi-peak sample, which is too small to make any robust conclusions. In the two lowest density bins, we have 16 and 26 HOP multi-peak galaxies, respectively.  Therefore, we may not have a large enough sample to identify the specific signal observed by Ellison et al. (2010).
 
There are also differences related to the method used to calculate local density.  We use all of the galaxies in each box to determine the projected local density for this rough comparison, but Ellison et al. (2010) choose galaxies within a line-of-sight velocity of 1000 km s$^{-1}$.  Also, we calculate local density using all the galaxies with M$_*$ $>$ 4 $\times$ 10$^9$ M$_\odot$ in our simulation, while Ellison et al. (2010) use only galaxies with M$_r$ $<$ -20.6.  If we were to only include galaxies with M$_r$ $<$ -20.6, we would eliminate 48\% of our C box galaxies and 60\% of our V box galaxies.  A direct comparison between our galaxies and those observed by Ellison et al. (2010) using the same projected local density bins is not robust, because we only have eleven paired galaxies in the lowest density bin used by Ellison et al. (2010) (log $\Sigma$ $<$ -0.55).  When we split our galaxy population into  three evenly log-spaced bins in projected local density (the same process as in Ellison et al. (2010)), we see the same trends as when we use the three-dimensional local density.   

\section{sSFR and Bound Pair Separation}\label{sec:interacting}

Another possible explanation for the differences in the bound pair (as defined in Section \ref{sec:bounddef}, a pair galaxy is gravitationally bound to at least one other galaxy that is not the cD of the largest cluster) and total populations is that bound galaxies are interacting.  Because even our closest pairs use a rather large distance bin (d $<$ 250 kpc $h^{-1}$), we could be diluting a signal that depends on the distance between galaxies.  Observations of close interacting pairs at low redshifts find that they are a blue population (e.g. Condon et al. 1982; Keel et al. 1985; Kennicutt et al. 1987; Wong et al. 2011; Patton et al. 2011), which is consistent with the blue galaxy population in our d $<$ 250 kpc $h^{-1}$ bound galaxies.  It is noteworthy that the HOP multi-peak sample is composed of strongly interacting galaxies, as we discussed earlier (Section \ref{sec:galaxyselection}, and have a larger blue population than the total galaxy population.  

\begin{figure*}
\includegraphics{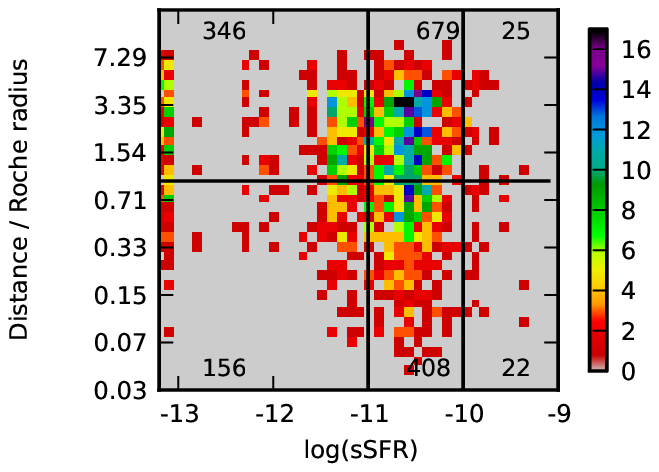}
\includegraphics{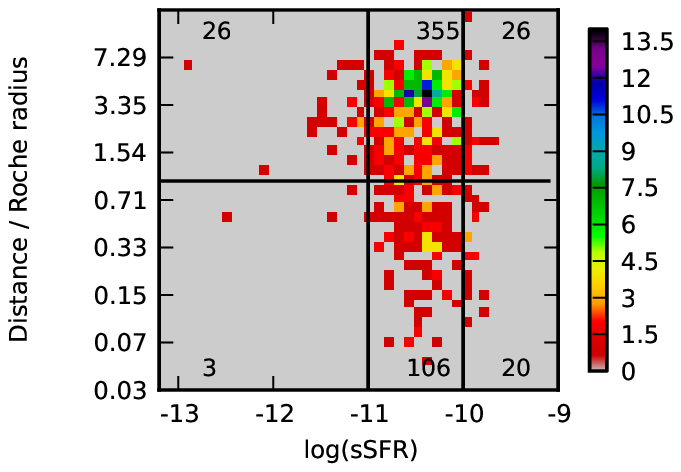}\\
\includegraphics{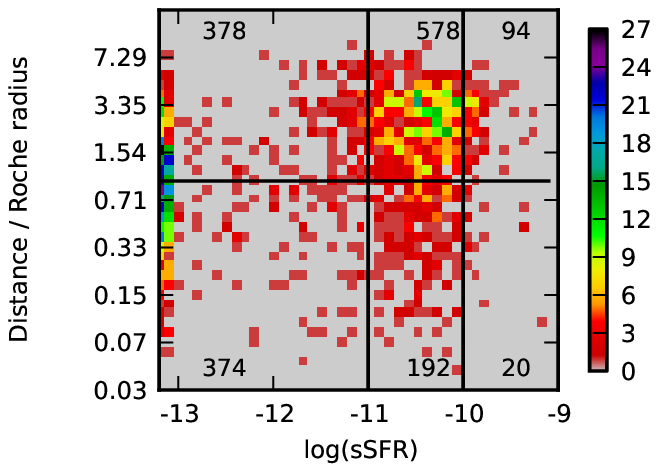}
\includegraphics{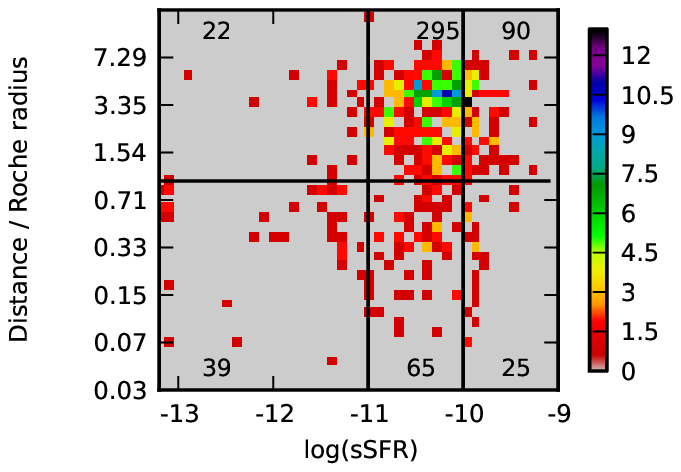}\\
\caption{Galaxy pair number in the (Distance/Roche radius) -- sSFR plane.  The top left panel is the major member of the pair in the C box, the bottom left panel is the minor member of the pair in the C box, the top right panel is the major member of the pair in the V box, and the bottom right panel is the minor member of the pair in the V box.  The sSFR histograms are evenly spaced in log space, except that the two bins below 10$^{-13}$ yr$^{-1}$ simply contain all the galaxies with sSFR below 10$^{-13}$ yr$^{-1}$.  In addition, we split each figure into six bins, separating non-star forming, star-forming, and highly star-forming galaxies that are inside or outside the Roche radius.  There is no strong trend of sSFR with galaxy distance.  There is a slight indication that the major galaxy in a pair may be more likely to be highly star-forming if the bound galaxy is within the Roche radius.}
\label{fig-interacting}
\end{figure*}

We can perform a quantitative check by noting that galaxies may be gravitationally strongly 
interacting if they are closer than the tidal distance, which we estimate as the Roche radius:
\begin{equation}
d = 160 h^{-1} kpc \times (\frac{M_{minor}}{10^{12} M_\odot})^{\frac{1}{3}} \times (2 \times (\frac{M_{major}}{M_{minor}}))^{\frac{1}{3}}.
\end{equation}
\noindent In Figure \ref{fig-interacting}, we plot the number density of pairs in the (Distance/Roche radius) - sSFR two-dimensional plane.  Note that the Roche radius is recalculated for every pair or group using the above equation.  The top left panel is the major member of the pair in the C box, the bottom left panel is the minor member of the pair in the C box, the top right panel is the major member of the pair in the V box, and the bottom right panel is the minor member of the pair in the V box.  The (Distance/Roche radius) bins are evenly spaced in log.  The sSFR bins are evenly spaced in log above 10$^{-13}$ yr$^{-1}$, and the two bins below 10$^{-13}$ yr$^{-1}$ contain all of the galaxies with sSFR lower than 10$^{-13}$ yr$^{-1}$.  We also split the x-axis into three separate groups:  non-star forming, star-forming, and starburst galaxies, and split the y-axis into two regions: inside or outside the Roche radius.  Each of these six bins is labeled with the number of galaxies it contains.  Examining the orbits of all of these bound galaxies is beyond the scope of this paper, but we do note that it is possible that galaxies that are currently outside of the Roche radius may have been closer at some point in the past.  

From these histograms, we see that there is no strong trend of sSFR with galaxy distance.  If we focus only on the larger bins, there is some indication that the major galaxy in a pair may be more likely to be a starburst galaxy if the minor bound galaxy is within the Roche radius, although the numbers of galaxies in the sSFR $>$ 10$^{-10}$ yr$^{-1}$ bins are small.  The major bound galaxies are slightly less likely to have low sSFR (sSFR $<$ 10$^{-11}$ yr$^{-1}$) if the minor galaxy is within its Roche radius.  These points could indicate that a major galaxy with the minor member of the pair within its Roche radius is able to accrete cold gas either from the minor member of the pair, or from gas in the nearby environment.  

The minor bound galaxy is more likely to have a sSFR $<$ 10$^{-11}$ yr$^{-1}$ if it is within the Roche radius of the major galaxy, and slightly less likely to have a sSFR $>$ 10$^{-10}$ yr$^{-1}$.  Thus, we do not see evidence of tidal triggering of star-formation in the smaller galaxy in a pair.  Instead, it seems that being bound to a larger galaxy decreases the sSFR.  We cannot say whether this is due to a gravitational interaction with the major member of the pair or if these galaxies are simply entering higher density environments, where any of a number of other mechanisms could redden the galaxy (e.g. harassment or ram pressure stripping, as discussed in the Introduction). 

Finally, if we consider all galaxies (major and minor) that are in pairs, we find that galaxies with the minor member inside the Roche radius are about 10\% less likely to be star-forming (sSFR $>$ 10$^{-11}$ yr$^{-1}$) than galaxies with the minor pair outside the Roche radius.  Therefore, although close pairs may be gravitationally interacting, we conclude that to first order these interactions are not resulting in the different sSFRs of close bound galaxies (d $<$ 250 $h^{-1}$ kpc plus HOP multi-peak galaxies) relative to the global galaxy population that we see in Figure \ref{fig-bluefrac}.

\section{A Coherent Physical Explanation}\label{sec:story}

On large scales, global structures, such as around a rich cluster of galaxies or a void region,
are expected to evolve differently, with the large-scale overdensity playing the role of a dynamic ``clock".
In this sense, a large-scale overdense region is dynamically more advanced than a large-scale underdense region.
This is in accord with one's intuitive expectation, for example, that a cluster is dynamically more developed
than a field region that is yet to collapse, which in turn is more advanced than a void region that may still
be experiencing expansion.  This macroscopic view is in line with the differences we see when comparing the C and V box total galaxy populations.  C box galaxies are more massive and have lower sSFRs and less cold gas than the V box population.  

It seems that local overdensity, as parameterized by $\rho_5$, plays a more important role with respect to star formation, galaxy colour, gas cooling, etc, at least at $z=0$ - $0.2$ range examined here.  We suggest that the local overdensity effectively determines the thermodynamic properties of the gas, primarily dictated by the strength of the converging shocks due to the collapse of embedding large-scale structures.  For example, in cluster environments the gas is heated to a high temperature commensurate with the depth of the potential well of the cluster, whereas in large-scale filaments and sheets the temperature is determined by the 2-d or 1-d Zel'dovich pancake collapse of the corresponding large-scale density fluctuations.
To zero-th order, a local density measure, such as $\rho_5$,
is a measure of the local gas temperature or perhaps more importantly the amount of cold gas (Cen 2011);
at a fixed galaxy halo mass, a higher $\rho_5$ corresponds to a lower amount of cold gas
than a lower $\rho_5$.  If we place this in the nomenclature of cold and hot mode accretion,
$\rho_5$ effectively measures the ``halo" mass of the embedding structure of the galaxy in question (although this is not necessarily limited to virialized systems).  There are also a number of interactions that can occur in large halos that may remove a galaxy's gas, as discussed in the Introduction.  Once a galaxy's gas has been exhausted or removed in these massive, hot halos, there is no simple mode for replenishing the reservoir (but see Kauffmann et al. 2010).

Within this context we can understand the trend of galaxies having more cold gas, and higher sSFRs
in low density regions than in high density regions for the general galaxy population.  We see the trend of increasing star-forming fraction with decreasing $\rho_5$ in Figure \ref{fig-bluefrac}.  
As we have shown, bound galaxy pairs tend to avoid cluster environments as well as void regions (Figures \ref{fig-pairfrac} and \ref{fig-pairfracld}).  As a result of this ``selection" effect, bound galaxy pairs tend to be redder than void galaxies but bluer than cluster galaxies.

However, this effect alone does not explain all the trends we have found for galaxy pairs.  Gravitational interactions between bound galaxies cause an additional, subtle physical effect that is 
dependent on local overdensity.  At a relatively high $\rho_5$, hot gas is more dominant in or surrounding a galaxy halo.
In this case, in our simulation close galaxy-galaxy interactions induce shocks, gas compression and cooling of hot gas that
causes a noticeable increase in the amount of cold gas, which in turn causes the interacting galaxies to 
have a higher star formation rate.
Thus, galaxy pairs in relatively high density regions tend to be bluer and more \ion{H}{I}-gas rich than the
general galaxy population at the same density. 
On the other hand, at a relatively low $\rho_5$ the amount of hot gas is negligible compared to the much larger amount of cold gas available.  Thus, any cooling of the hot halo caused by close galaxy-galaxy interactions does not substantially increase the amount of cold gas.  
Indeed, some of the initially cold gas could be shock-heated to hot gas or disrupted during the interactions.
Consequently, galaxy pairs in relatively low density regions tend to have colours and \ion{H}{I}-fractions similar to the total population in those regions.  

Whether the cooling of hot halo gas influences the sSFR of galaxies depends on whether the galaxies at that local density are rich or poor in \ion{H}{I}.  In our pair galaxies, the replacement of direct cold gas accretion with cooling gas from the hot halo can be most directly seen by examining Figure \ref{fig-mhifrac}.  The \ion{H}{I} fraction decreases as the local density increases, and only at the highest local densities do bound galaxies have higher \ion{H}{I} fractions than the comparison population.  This cooling from the hot halo is a smaller effect than the strong dependence of \ion{H}{I} gas content on local density, as evidenced by the smaller difference between the fraction of \ion{H}{I}-rich bound and matched samples in the highest density bin compared to the difference between the ``total" populations in the middle and highest density bins.

The properties of close unbound galaxies support this scenario.  They have redder colours and lower \ion{H}{I} fractions than the bound galaxies, which are undergoing a more extended interaction.  Also, unbound close galaxies are in higher density environments, and because of the strong correlation we find between local density and galaxy colour we would expect them to be redder.

\section{Conclusion}\label{sec:conclusion}

We have examined the galaxies formed in a large, 120 $h^{-1}$ Mpc cosmological 
simulation in detail in two zoomed-in boxes:  
one centred on a void of size 31 $\times$ 31 $\times$ 35 $h^{-3}$ Mpc$^3$
and the other on a cluster of size  21 $\times$  24 $\times$ 20 $h^{-3}$ Mpc$^3$. 
The two refined regions have an overlapping range of local densities, which allows us to separate local and large-scale environmental effects.  We have utilized the high resolution (0.46 $h^{-1}$ kpc) in these regions to examine galaxies in detail to determine how being a member of a pair (as defined in Section \ref{sec:bounddef}, a pair galaxy is gravitationally bound to at least one other galaxy that is not the cD of the largest cluster) affects galaxy properties and whether these effects depend on environment.  We now summarize our results and conclusions.

1)  We use the three-dimensional information provided by our simulation to determine whether the observational criteria used to choose projected pairs identify physically bound pairs.  We find that these criteria yield pair samples that include a larger fraction of spurious pairs than in previous work (Perez et al. 2006a found 27\% spurious fraction to our 48\%).  If we only consider projected pairs in the lower-density V box, we find a much lower spurious pair fraction of 12\%.  In qualitative agreement with previous work, we find that the fraction of spurious pairs increases in dense regions.  In order to agree with Perez et al. (2006a), we must only consider regions with $\rho_5$ $\le$ 1 $h^3$ Mpc$^{-3}$, which is less dense than group environments.  In order to understand if pair interactions may redden galaxies, it is important to consider higher local densities where a larger fraction of the galaxy population is red.  The high fraction of spurious pairs at high densities (58\% at $\rho_5$ $>$ 1 $h^3$ Mpc$^{-3}$) may have masked the trends we find for pair galaxies as a function of local density in some observed samples.

2)  We find that, of all pairs, (59\%,13\%, 28\%)  are wet, dry and mixed pairs, respectively, which are in excellent agreement with the corresponding fractions of (56\%, 15\%, 29\%) found in observed galaxy pairs by Lin et al. (2010).   In Figure \ref{fig-wdmhists} we show that the environments of these pairs are consistent with the colour-density relation.



3) We see no evidence that, considering the bound population as a whole, being closer than the Roche radius enhances the sSFR of gravitationally bound pair galaxies (Section \ref{sec:interacting}).

4) Largely independent of being in a galaxy pair or not, the strongest trend we find is environmental, where the average sSFR is a monotonic increasing function of local environmental density, defined by $\rho_5$.  This trend is in broad agreement with observations from SDSS. The likely physical cause for this trend is the underlying trend of higher gas temperature  in higher density regions due to formation of massive halos or large scales structures such as pancakes and filaments (Kere{\v s} et al. 2005, Dekel \& Birnboim 2006; Cen 2011), which dictates the cold gas supply.  This may also have to do with the many interactions that can occur in high density regions:  galaxy-cluster, galaxy-galaxy, and galaxy-intracluster medium interactions can all remove galaxy's cold gas (e.g. Merritt 1984; Moore et al. 1996; Gunn \& Gott 1972; Larson et al. 1980).  

5) Being a pair galaxy has a secondary but significant effect on star formation rate and cold gas on pair galaxies. We find that 
galaxies in close pairs tend to have higher sSFR in high density environment than the non-paired galaxies at the same environment density (e.g., in and near galaxy clusters).  Such an enhancement for pairs is not seen in low density environments
(e.g., in the centre of a void); in fact, the opposite is seen there.  As a result, we find that pairs in high density environments tend to have fewer low sSFR galaxies than non-pairs, where pairs in low density environments have the opposite trend.
However, very close pairs (r$_p$ $<$ 50 kpc, the HOP multi-peak sample) always have an elevated fraction of high sSFR galaxies, regardless of their environment.  Analogous statements apply to \ion{H}{I} gas in galaxies.  

6) Interactions between close pairs affects galaxies at high local densities by cooling gas out of the hot halo, as evidenced by the higher M$_{HI}$/M$_*$ of close pairs in higher local density bins rather than in lower local density bins where cold flow accretion is more common and hot gas is largely absent (Figure \ref{fig-mhifrac} and Section \ref{sec:story}).  

Our results indicate that observationally finding galaxy pairs without resorting to searches for tidally-disturbed galaxies can be difficult.  That said, we do find similar results--a general increase in sSFR as galaxy pairs move closer together.  We predict that if observers look for pairs specifically in and near voids, they will find that in these large-scale underdensities, the pair population will not have a larger fraction of star-forming galaxies than the total void population (although we do see hints that the strongly star-forming fraction may be increased in pairs in voids in Figure \ref{fig-verybluefrac}).  

Finally, we introduced this paper by claiming that mergers may increase the mass of star-forming galaxies, drive the evolution of star-forming galaxies into red and dead galaxies, and hierarchically build massive red galaxies through dry mergers.  Do our results and the physical explanation we put forth in Section \ref{sec:story} support that claim? The answer appears to be no, as we will now discuss.

First, we find that the majority of our close pairs (d $<$ 250 $h^{-1}$ kpc galaxies plus d $<$ 50 $h^{-1}$ kpc galaxies--the HOP multi-peak galaxies) are star-forming.  If we assume that these close pairs will merge, this indicates that close pair interaction of galaxies do not quench star formation prior to the merger.  The fraction of strongly star-forming galaxies is very similar between the close bound pairs and total galaxy populations, indicating no significantly strong bursts of star-formation prior to merging.  Second, the fuel for star formation (M$_{HI}$/M$_{*}$) increases as the separation of the galaxy pair decreases.  This suggests that at least during the pair interaction period prior to the final merger the  gas reservoir is increased rather than decreased, even though star formation rate is higher for closer pairs.  In other words, there is no sign of pair interactions causing cold gas exhaustion.  Third, although the close pairs (d $<$ 250 kpc $h^{-1}$ plus galaxies with separations of 30-50 kpc $h^{-1}$--the HOP multi-peak sample) have some of the highest sSFRs of the entire galaxy population, they only account for 15 of the 39 highest sSFRs.  This shows that pairs (including very close  pairs) do not monopolize starbursts. Could the remaining (24 out of 39) highest sSFR non-pair galaxies 
be post-merger galaxies?  We find that of the 17 galaxies with sSFR $\ge$ 3 $\times$ 10$^{-10}$ yr$^{-1}$ that we can track back at least one output (galaxies at z $\le$ 0.15 and that had M$_*$ $>$ 10$^{9.6}$ M$_{\odot}$ in the previous output), only one galaxy is post-merger (had been a member of a close pair the output before, but is no longer a member of a pair).  Altogether, this evidence suggests that in general neither mergers nor pre-merger interactions quench star formation either through heating cold gas or a burst of star formation.  The $\sim$40\% (15/39) of starbursting galaxies that are in pairs falls in the range of poststarburst galaxies observed to be merging or tidally-interacting:  about 50\% in Poggianti et al. (1999) and 25\% in Zabludoff et al. (1996).

In fact, rather than pair interactions driving a starburst that exhausts a galaxy's reservoir of cold gas, we find that close pairs (d $<$ 250 $h^{-1}$ kpc galaxies plus galaxies with separations of 30-50 kpc $h^{-1}$--the HOP multi-peak sample) have a higher (M$_{HI}$/M$_{*}$).  The higher star-forming fraction of close pairs, particularly at higher local densities (and therefore in more massive halos) may suggest that the small amount of star formation and \ion{H}{I} observed in otherwise red and dead disturbed galaxies is created from cooling of hot halo gas in interacting elliptical galaxies (Donovan et al. 2007), and therefore ``dry" mergers may hierarchically grow larger red galaxies as well as induce small amounts of star formation.  Donovan et al. (2007) frequently find a peculiar morphology in the \ion{H}{I} associated with their observed red elliptical galaxies, possibly indicating a recent tidal disturbance. 

Are mergers responsible for the colour-density relation?  Available evidence suggest the answer is no.
We see no evidence of close pair interactions
causing accelerated removal of cold gas reservoir.
Nor do we see evidence of pair interactions causing star formation to cease
in the outskirts of clusters of galaxies, before their entry into clusters.
All evidence points to the opposite of what the colour-density relation suggests.
Nevertheless, our galaxies are in broad agreement with the observed
colour-density relation, which strongly suggests that it is the environment,
rather than pair-interactions/mergers, that is responsible for the colour-density relation. This could be due to any of several interactions that can occur in high-density environments:  galaxy-cluster, galaxy-galaxy, and galaxy-intracluster medium interactions can all remove galaxy's cold gas (e.g. Merritt 1984; Moore et al. 1996; Gunn \& Gott 1972).  Also, residing in a massive halo means that galaxies do not have access to cold flows that can replenish their gas reservoir, so they will eventually be unable to form stars (Larson et al. 1980; Kere{\v s} et al. 2005).  We shall devote a more detailed investigation to this important issue.


\vskip 1cm
We would like to thank Dr. M.K.R. Joung for help on
generating initial conditions for the simulations and running a portion
of the simulations and Greg Bryan for help with Enzo code.  ST would like to that Jacqueline van Gorkom for helpful comments on this paper.  We would like to thank our referee for comments that greatly improved the quality of the paper.
Computing resources were in part provided by the NASA High-
End Computing (HEC) Program through the NASA Advanced
Supercomputing (NAS) Division at Ames Research Center.
This work is supported in part by grant NNX11AI23G.

\appendix

\section{The Redshift Distribution}\label{sec:redshiftexplain}

We have shown that there is a higher fraction of bound pair galaxies at higher redshifts (Section \ref{sec:redshift} and Table \ref{tbl-pairs}).  Because at higher redshifts our galaxy population tends to be bluer and more \ion{H}{I}-rich, our results could partially reflect this difference in the populations.  

In order to determine if this is the case, we plot each set of CDFs for each individual redshift output.  We find that the results at any single redshift are consistent with the results combining all the redshifts, so the differences in the redshift distributions of the total galaxy sample and the bound galaxy samples should not be the main cause of these interesting trends.  

We can go through each CDF in a bit more detail.  First, when we examine the sSFR we find that at all redshifts, the C box bound galaxies have a smaller fraction of low sSFR galaxies than the corresponding total C population (as in the combined redshifts shown in Figure \ref{fig-ssfr}).  In the V box, at all redshifts the total pair and close pair (d $<$ 250 $h^{-1}$ kpc) populations have a slightly higher fraction of low sSFR galaxies.  These separated redshift results agree with the combined redshift results shown in Figure \ref{fig-ssfr}.  There is a higher fraction of pair galaxies with high sSFRs at z=0.15 and z=0.2, while the fraction of high sSFR galaxies is very similar in the pair and total V populations at z = 0 and z = 0.05.  However, z=0.05 has an excess in the blue population of HOP multi-peak galaxies in both the C and V boxes.  At any redshift pairs either have a higher fraction of high sSFR galaxies (z = 0.05, 0.15, 0.2) or have the same fraction of high sSFR galaxies compared to the total population (z = 0).  

Comparing to Figure \ref{fig-smhi}, in the \ion{H}{I}-fraction CDFs we find that  at all redshifts, the C box bound galaxies have a smaller fraction of \ion{H}{I}-poor galaxies than the total C population, and the fraction decreases when we consider only close bound pairs (d $<$ 250 $h^{-1}$ kpc).  In the V box, at all redshifts the fraction of \ion{H}{I}-poor galaxies is similar between bound galaxies and the total populations.  Again, these trends agree with the combined redshifts shown in Figure \ref{fig-smhi}.  At all redshifts above z=0 V pairs have a higher fraction of \ion{H}{I}-rich galaxies than the total V population.  Finally, at all redshifts but z = 0 the C bound galaxies are at higher local densities than the total populations.  In the V box, at all redshifts pairs tend to be at higher local densities than the total V galaxy population.  These trends agree with the trends in Figure \ref{fig-localdens}.

In addition, we compare the star-forming fraction of pairs and the total population for each individual redshift output.  As in the CDF comparisons, at each individual redshift the results are consistent with Figure \ref{fig-bluefrac}.  Even if we plot the star-forming fraction of close pairs (d $<$ 250 kpc $h^{-1}$ plus HOP multi-peak) from all redshift outputs against the star-forming fraction of only the z = 0.2 total population we find nearly identical results.  When comparing close bound pairs to the total galaxy population at z = 0.2 only, in the two lower density bins the total population has a slightly higher star-forming fraction than the bound population, in the third local density bin the star-forming fraction of close pairs is higher than that of the total population, and in the fourth local density bin the star-forming fraction of close pairs is higher than that of the total population with large, overlapping error bars.

It is possible that because we observe many of our galaxies multiple times, the pair population happens to be bluer and more \ion{H}{I}-rich than the total population throughout the simulation for reasons other than being a member of a bound pair.  Although we must admit this is a possibility, we do not consider it very likely.  Indeed, we can think of no other commonality that may be responsible for these galaxy properties.  We conclude that the redshift distribution of the galaxies is not the main driver of our results.

\section{Comparison with Close Unbound galaxies}\label{sec:unbound} 

\begin{figure}
\includegraphics{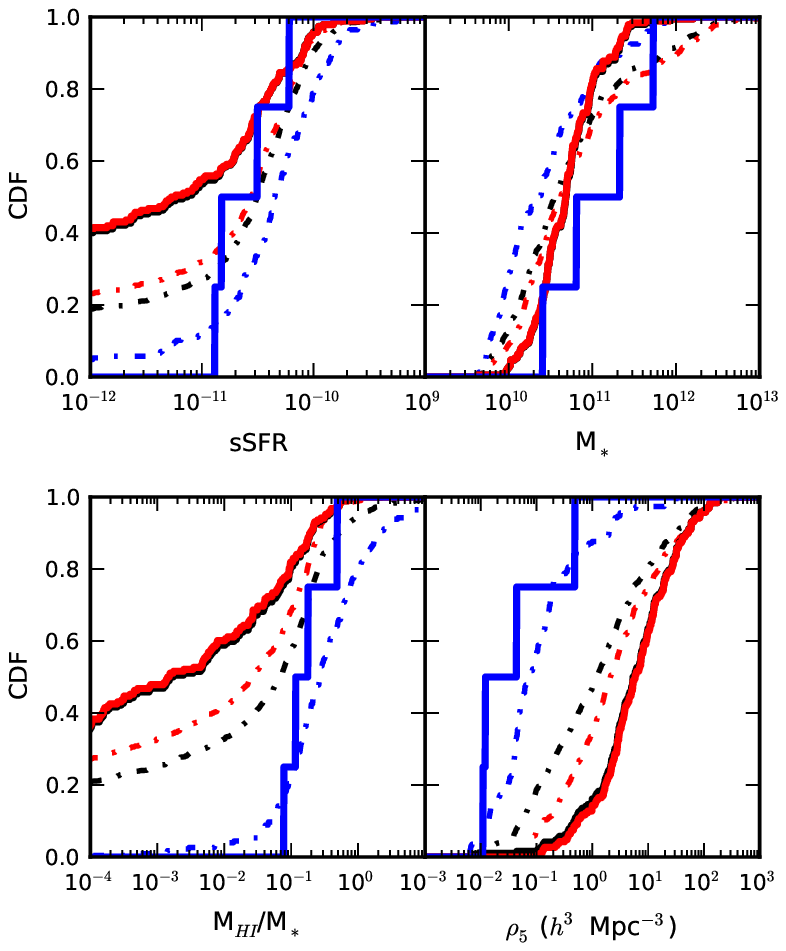}
\caption{The CDFs for close but \textit{unbound} galaxies (solid curves).  For comparison, each panel also shows the CDFs of the close bound galaxies (d $<$ 250 kpc $h^{-1}$) as dash-dotted lines.  See Section \ref{sec:unbound} for discussion.}  
\label{fig-pairnopair}
\end{figure}

One advantage of using a simulation is that we can determine whether close galaxies are in fact gravitationally bound.  We can determine whether being in a bound pair (as defined in Section \ref{sec:bounddef}, a pair galaxy is gravitationally bound to at least one other galaxy that is not the cD of the largest cluster) affects a galaxy differently than simply being in the midst of a close fly-by.  In order to do this, we compare close bound galaxies (d $<$ 250 $h^{-1}$ kpc) to galaxies that are not bound to any other galaxy but are within 250 $h^{-1}$ kpc of another galaxy, shown in Figure \ref{fig-pairnopair}.  Because the samples are small, we do not attempt to match the mass distribution of the bound galaxies with d $<$ 250 kpc $h^{-1}$ and include a CDF of M$_*$.  We also include CDFs of $\rho_5$ in this figure for uniformity, but we will discuss this panel in later sections.  In Figure \ref{fig-pairnopair}, the dashed lines are the pairs within d $<$ 250 kpc $h^{-1}$, and the solid lines are close unbound galaxies.  Although we will discuss the C box (red lines) and V box (blue lines) galaxies separately, we also include the C plus V box samples in the Figure as the black lines.  As usual, there are more C box galaxies, and they dominate the C plus V box samples to the point that the black lines can barely be seen for the unbound close galaxies.

We first consider the C box (red lines).  Even though we choose not to include any close unbound galaxies within 1.4 $h^{-1}$ Mpc of the cluster centre so as not to skew the results with red cluster galaxies, being in a bound pair results in different galaxy properties than simply having a close fly-by.  Bound galaxies have a higher fraction of galaxies with sSFR $>$ 10$^{-11}$ yr$^{-1}$ than the close but unbound galaxies.  Similarly, close unbound galaxies tend to have lower \ion{H}{I} gas fractions than close bound galaxies, although the overall ranges of M$_{HI}$/M$_*$ are similar.  The lower sSFRs and \ion{H}{I} fractions of the unbound galaxies cannot be attributed to the difference in the underlying stellar mass distributions, because the unbound galaxies tend to be less massive than close bound pairs.

Now, we turn to V box galaxies.  We see that the sSFR distribution of unbound galaxies spans a smaller range in sSFR than that of the bound galaxies.  Considering that there are only 4 galaxies in the unbound sample, we cannot conclude that this is or is not a close match to the bound sample.  The \ion{H}{I} mass-fraction CDFs compare similarly to the sSFR CDFs.   In contrast to the C box, the unbound galaxies in the V box are on the most massive end of the V bound galaxy distribution, although the small number of unbound galaxies makes it difficult to draw firm conclusions.

It is likely that the M$_*$ (upper right panel) and $\rho_5$ (lower right panel) CDFs for the C box close unbound galaxies are affected by the exclusion of galaxies within 1.4 $h^{-1}$ Mpc of the cluster centre.  Despite that, the common trends for both V and C boxes are that close unbound galaxies tend to be redder and less \ion{H}{I} rich than bound pairs.  Including cluster galaxies would likely strengthen these trends.

\section{Median sSFR of Star-Forming Galaxies}\label{sec:medianssfr}

We also examine whether the environment or pair membership affects the sSFR of star-forming galaxies (sSFR $>$ 10$^{-11}$ yr$^{-1}$).  We find that the median sSFR of star-forming galaxies increases by a factor of two from our highest to our lowest local density, but with very large error bars.  There is no discernible difference between the V box and C box galaxies, except perhaps in the highest density bin, where the star-forming V box galaxies have a marginally higher median sSFR.  Being a member of a pair has a negligible affect on the median sSFR of star-forming galaxies.  

\section{Star-Forming Fraction Relative to Galaxy Mass}\label{sec:ssfrmass}

As with all of our results, we have to consider whether the decrease in star-forming fraction is due to local density or to the fact that the mass of galaxies increases with local density.  We have tested this in a few ways.  First, we split our data set into bins of stellar mass instead of local density.  We find that the star-forming fraction is larger by more than 15\% in galaxies with masses below 3 $\times$ 10$^{10}$ M$_{\odot}$ than in galaxies with stellar masses above 3 $\times$ 10$^{10}$ M$_{\odot}$.  This variation with mass is much smaller than the variation we find with local density.  Second, we considered galaxies only within a certain stellar mass range:  below 3 $\times$ 10$^{10}$ M$_{\odot}$, above 3 $\times$ 10$^{10}$ M$_{\odot}$, and above 10$^{12}$ M$_{\odot}$.  Using the same local density bins, the galaxies above and below 3 $\times$ 10$^{10}$ M$_{\odot}$ have very similar star-forming fractions, with the lower mass galaxies having slightly higher star-forming fraction at the lower densities ($\le$ 7\% higher fraction), and a slightly lower star-forming fraction in the highest density bin (a 3\% smaller star-forming fraction).  If we only consider our extremely high mass galaxies, with stellar masses above 10$^{12}$ M$_{\odot}$, we can only look at the two higher local density bins, and the fractions are the same to well within the error bars (and with a star-forming fraction that is nearly 25\% above the star-forming fraction of the lower mass galaxies in the highest local density bin).  Therefore, we conclude that this trend of star formation with local density is mostly due to local density, not to galaxy mass.

As we have discussed (for example, in Section \ref{sec:galaxyselection}), although our high-mass galaxies are redder than our low-mass galaxies, they have more star formation than observed galaxies, and some continue to grow to high stellar masses.  This is likely part of the reason that we see little difference in the star-forming fraction of galaxies at different masses.  This problem does not affect our results because they are based on comparisons within our simulation.  If our star-forming fraction is high, it will be high in both bound and unbound galaxies.  If anything, the higher SFR for some large galaxies (cD, etc) will weaken the environmental trends.  

\section{Star-Forming Fraction Relative to the Local Density Distribution}\label{sec:ssfrlocald}

Because our result that being a member of a bound pair boosts (by more than 10\%) the star-forming fraction of the galaxy population only at the highest local densities is an important result, we consider the fact that even in each of our four local density bins close bound pair galaxies may have a different $\rho_5$ distribution than the total population.  Therefore, in addition to mass-matching, we also match the CDFs of $\rho_5$.  Similar to before, we find that in the two lower local density bins the bound and matched samples' star-forming fractions are the same within the error bars, and in the two highest local density bins the bound galaxies have a significantly higher star-forming fraction than the matched (in both mass and local density) population, the same trend as without $\rho_5$ and mass matching.

\label{lastpage}


\begin{thebibliography}{99}
\bibitem[\protect\citeauthoryear{Alonso et al.}{2006}]{2006RMxAC..26..187A} 
Alonso M.~S., Tissera P.~B., Lambas D.~G., Coldwell G., 2006, RMxAC, 26, 
187 


\bibitem[\protect\citeauthoryear{Alonso et al.}{2004}]{2004MNRAS.352.1081A} 
Alonso M.~S., Tissera P.~B., Coldwell G., Lambas D.~G., 2004, MNRAS, 352, 
1081 


\bibitem[\protect\citeauthoryear{Baldry et al.}{2004}]{2004ApJ...600..681B} 
Baldry I.~K., Glazebrook K., Brinkmann J., Ivezi{\'c} {\v Z}., Lupton 
R.~H., Nichol R.~C., Szalay A.~S., 2004, ApJ, 600, 681 


\bibitem[\protect\citeauthoryear{Balogh et al.}{2001}]{2001MNRAS.326.1228B} 
Balogh M.~L., Pearce F.~R., Bower R.~G., Kay S.~T., 2001, MNRAS, 326, 1228 


\bibitem[\protect\citeauthoryear{Barnes 
\& Hernquist}{1996}]{1996ApJ...471..115B} Barnes J.~E., Hernquist L., 1996, ApJ, 471, 115 


\bibitem[\protect\citeauthoryear{Barton et al.}{2007}]{2007ApJ...671.1538B} 
Barton E.~J., Arnold J.~A., Zentner A.~R., Bullock J.~S., Wechsler R.~H., 
2007, ApJ, 671, 1538 


\bibitem[\protect\citeauthoryear{Bell et al.}{2004}]{2004ApJ...608..752B} 
Bell E.~F., et al., 2004, ApJ, 608, 752 


\bibitem[\protect\citeauthoryear{Berrier et 
al.}{2009}]{2009ApJ...690.1292B} Berrier J.~C., Stewart K.~R., Bullock 
J.~S., Purcell C.~W., Barton E.~J., Wechsler R.~H., 2009, ApJ, 690, 1292 


\bibitem[\protect\citeauthoryear{Blanton}{2006}]{2006ApJ...648..268B} 
Blanton M.~R., 2006, ApJ, 648, 268 


\bibitem[\protect\citeauthoryear{Blanton et 
al.}{2003}]{2003ApJ...594..186B} Blanton M.~R., et al., 2003, ApJ, 594, 186 


\bibitem[\protect\citeauthoryear{Bruzual 
\& Charlot}{2003}]{2003MNRAS.344.1000B} Bruzual G., Charlot S., 2003, MNRAS, 344, 1000 


\bibitem[\protect\citeauthoryear{Bryan}{1999}]{1999CoScE...1...46B} Bryan 
G.~L., 1999, CoScE, 1, 46 


\bibitem[\protect\citeauthoryear{Butcher 
\& Oemler}{1978}]{1978ApJ...226..559B} Butcher H., Oemler A., Jr., 1978, ApJ, 226, 559 


\bibitem[\protect\citeauthoryear{Byrd 
\& Valtonen}{1990}]{1990ApJ...350...89B} Byrd G., Valtonen M., 1990, ApJ, 350, 89 


\bibitem[\protect\citeauthoryear{Cen}{2011}]{2011ApJ...741...99C} Cen R., 
2011, ApJ, 741, 99 


\bibitem[\protect\citeauthoryear{Cen}{2010}]{2010arXiv1010.5014C} Cen R., 
2010, arXiv, arXiv:1010.5014 


\bibitem[\protect\citeauthoryear{Cen et al.}{1995}]{1995ApJ...451..436C} 
Cen R., Kang H., Ostriker J.~P., Ryu D., 1995, ApJ, 451, 436 


\bibitem[\protect\citeauthoryear{Cen, Nagamine, 
\& Ostriker}{2005}]{2005ApJ...635...86C} Cen R., Nagamine K., Ostriker J.~P., 2005, ApJ, 635, 86 


\bibitem[\protect\citeauthoryear{Cen 
\& Ostriker}{1992}]{1992ApJ...399L.113C} Cen R., Ostriker J.~P., 1992, ApJ, 399, L113 


\bibitem[\protect\citeauthoryear{Chandrasekhar}{1961}]{1961hhs..book.....C} 
Chandrasekhar S., 1961, Hydrodynamic and Hydromagnetic Stability. International Series of Monographs on Physics, Oxford: Clarendon


\bibitem[\protect\citeauthoryear{Condon et al.}{1982}]{1982ApJ...252..102C} 
Condon J.~J., Condon M.~A., Gisler G., Puschell J.~J., 1982, ApJ, 252, 102 


\bibitem[\protect\citeauthoryear{Cooper et al.}{2008}]{2008MNRAS.383.1058C} 
Cooper M.~C., et al., 2008, MNRAS, 383, 1058 


\bibitem[\protect\citeauthoryear{Cowie 
\& Songaila}{1977}]{1977Natur.266..501C} Cowie L.~L., Songaila A., 1977, Natur, 266, 501 


\bibitem[\protect\citeauthoryear{Darg et al.}{2010}]{2010MNRAS.401.1552D} 
Darg D.~W., et al., 2010, MNRAS, 401, 1552 


\bibitem[\protect\citeauthoryear{De Lucia et 
al.}{2012}]{2012MNRAS.tmp.3042D} De Lucia G., Weinmann S., Poggianti B.~M., 
Arag{\'o}n-Salamanca A., Zaritsky D., 2012, MNRAS, 3042 


\bibitem[\protect\citeauthoryear{Dekel 
\& Birnboim}{2006}]{2006MNRAS.368....2D} Dekel A., Birnboim Y., 2006, MNRAS, 368, 2 


\bibitem[\protect\citeauthoryear{Donovan, Hibbard, 
\& van Gorkom}{2007}]{2007AJ....134.1118D} Donovan J.~L., Hibbard J.~E., van Gorkom J.~H., 2007, AJ, 134, 1118 


\bibitem[\protect\citeauthoryear{Dressler}{1980}]{1980ApJ...236..351D} 
Dressler A., 1980, ApJ, 236, 351 


\bibitem[\protect\citeauthoryear{Eisenstein 
\& Hut}{1998}]{1998ApJ...498..137E} Eisenstein D.~J., Hut P., 1998, ApJ, 498, 137 


\bibitem[\protect\citeauthoryear{Ellison et 
al.}{2008}]{2008AJ....135.1877E} Ellison S.~L., Patton D.~R., Simard L., 
McConnachie A.~W., 2008, AJ, 135, 1877 


\bibitem[\protect\citeauthoryear{Ellison et 
al.}{2010}]{2010MNRAS.407.1514E} Ellison S.~L., Patton D.~R., Simard L., 
McConnachie A.~W., Baldry I.~K., Mendel J.~T., 2010, MNRAS, 407, 1514 


\bibitem[\protect\citeauthoryear{Faber et al.}{2007}]{2007ApJ...665..265F} 
Faber S.~M., et al., 2007, ApJ, 665, 265 


\bibitem[\protect\citeauthoryear{G{\'o}mez et 
al.}{2003}]{2003ApJ...584..210G} G{\'o}mez P.~L., et al., 2003, ApJ, 584, 
210 


\bibitem[\protect\citeauthoryear{Gunn 
\& Gott}{1972}]{1972ApJ...176....1G} Gunn J.~E., Gott J.~R., III, 1972, ApJ, 176, 1 


\bibitem[\protect\citeauthoryear{Haardt 
\& Madau}{1996}]{1996ApJ...461...20H} Haardt F., Madau P., 1996, ApJ, 461, 20 


\bibitem[\protect\citeauthoryear{Haynes, Giovanelli, 
\& Chincarini}{1984}]{1984ARA&A..22..445H} Haynes M.~P., Giovanelli R., Chincarini G.~L., 1984, ARA\&A, 22, 445 


\bibitem[\protect\citeauthoryear{Heinis et al.}{2009}]{2009ApJ...698.1838H} 
Heinis S., et al., 2009, ApJ, 698, 1838 


\bibitem[\protect\citeauthoryear{Hibbard 
\& van Gorkom}{1996}]{1996AJ....111..655H} Hibbard J.~E., van Gorkom J.~H., 1996, AJ, 111, 655 


\bibitem[\protect\citeauthoryear{Hubble 
\& Humason}{1931}]{1931ApJ....74...43H} Hubble E., Humason M.~L., 1931, ApJ, 74, 43 


\bibitem[\protect\citeauthoryear{Joung, Cen, 
\& Bryan}{2009}]{2009ApJ...692L...1J} Joung M.~R., Cen R., Bryan G.~L., 2009, ApJ, 692, L1 


\bibitem[\protect\citeauthoryear{Kauffmann et 
al.}{2004}]{2004MNRAS.353..713K} Kauffmann G., White S.~D.~M., Heckman 
T.~M., M{\'e}nard B., Brinchmann J., Charlot S., Tremonti C., Brinkmann J., 
2004, MNRAS, 353, 713 

\bibitem[\protect\citeauthoryear{Kauffmann, Li, 
\& Heckman}{2010}]{2010MNRAS.409..491K} Kauffmann G., Li C., Heckman T.~M., 2010, MNRAS, 409, 491 

\bibitem[\protect\citeauthoryear{Keel et al.}{1985}]{1985AJ.....90..708K} 
Keel W.~C., Kennicutt R.~C., Jr., Hummel E., van der Hulst J.~M., 1985, AJ, 
90, 708 


\bibitem[\protect\citeauthoryear{Kennicutt et 
al.}{1987}]{1987AJ.....93.1011K} Kennicutt R.~C., Jr., Roettiger K.~A., 
Keel W.~C., van der Hulst J.~M., Hummel E., 1987, AJ, 93, 1011 


\bibitem[\protect\citeauthoryear{Kere{\v s} et 
al.}{2005}]{2005MNRAS.363....2K} Kere{\v s} D., Katz N., Weinberg D.~H., 
Dav{\'e} R., 2005, MNRAS, 363, 2 


\bibitem[\protect\citeauthoryear{Kewley, Geller, 
\& Barton}{2006}]{2006AJ....131.2004K} Kewley L.~J., Geller M.~J., Barton E.~J., 2006, AJ, 131, 2004 


\bibitem[\protect\citeauthoryear{Kewley et al.}{2010}]{2010ApJ...721L..48K} 
Kewley L.~J., Rupke D., Zahid H.~J., Geller M.~J., Barton E.~J., 2010, ApJ, 
721, L48 


\bibitem[\protect\citeauthoryear{Komatsu et 
al.}{2011}]{2011ApJS..192...18K} Komatsu E., et al., 2011, ApJS, 192, 18 


\bibitem[\protect\citeauthoryear{Kreckel, Joung, 
\& Cen}{2011}]{2011ApJ...735..132K} Kreckel K., Joung M.~R., Cen R., 2011, ApJ, 735, 132 


\bibitem[\protect\citeauthoryear{Lambas et al.}{2003}]{2003MNRAS.346.1189L} 
Lambas D.~G., Tissera P.~B., Alonso M.~S., Coldwell G., 2003, MNRAS, 346, 
1189 


\bibitem[\protect\citeauthoryear{Larson 
\& Tinsley}{1978}]{1978ApJ...219...46L} Larson R.~B., Tinsley B.~M., 1978, ApJ, 219, 46 


\bibitem[\protect\citeauthoryear{Larson, Tinsley, 
\& Caldwell}{1980}]{1980ApJ...237..692L} Larson R.~B., Tinsley B.~M., Caldwell C.~N., 1980, ApJ, 237, 692 


\bibitem[\protect\citeauthoryear{Lin et al.}{2008}]{2008ASPC..399..298L} 
Lin L., Patton D.~R., Koo C.~D., Casteels K., Hsieh B.~C., 2008, ASPC, 399, 
298 


\bibitem[\protect\citeauthoryear{Lin et al.}{2010}]{2010ApJ...718.1158L} 
Lin L., et al., 2010, ApJ, 718, 1158 


\bibitem[\protect\citeauthoryear{Makino 
\& Hut}{1997}]{1997ApJ...481...83M} Makino J., Hut P., 1997, ApJ, 481, 83 


\bibitem[\protect\citeauthoryear{Mamon}{1986}]{1986ApJ...307..426M} Mamon 
G.~A., 1986, ApJ, 307, 426 


\bibitem[\protect\citeauthoryear{Martin et al.}{2007}]{2007ApJS..173..342M} 
Martin D.~C., et al., 2007, ApJS, 173, 342 


\bibitem[\protect\citeauthoryear{McGee et al.}{2009}]{2009MNRAS.400..937M} 
McGee S.~L., Balogh M.~L., Bower R.~G., Font A.~S., McCarthy I.~G., 2009, 
MNRAS, 400, 937 


\bibitem[\protect\citeauthoryear{Merritt}{1984}]{1984ApJ...276...26M} 
Merritt D., 1984, ApJ, 276, 26 


\bibitem[\protect\citeauthoryear{Mihos}{2004}]{2004cgpc.symp..277M} Mihos 
J.~C., 2004, in J.S. Mulchaey, A. Dressler, and A. Oemler, eds, Clusters of Galaxies: Probes of Cosmological Structure and Galaxy Evolution. Cambridge University Press, p. 277.


\bibitem[\protect\citeauthoryear{Moore et al.}{1996}]{1996Natur.379..613M} 
Moore B., Katz N., Lake G., Dressler A., Oemler A., 1996, Natur, 379, 613 


\bibitem[\protect\citeauthoryear{Moss}{2006}]{2006MNRAS.373..167M} Moss C., 
2006, MNRAS, 373, 167 


\bibitem[\protect\citeauthoryear{Nikolic, Cullen, 
\& Alexander}{2004}]{2004MNRAS.355..874N} Nikolic B., Cullen H., Alexander P., 2004, MNRAS, 355, 874 


\bibitem[\protect\citeauthoryear{Nulsen}{1982}]{1982MNRAS.198.1007N} Nulsen 
P.~E.~J., 1982, MNRAS, 198, 1007 


\bibitem[\protect\citeauthoryear{O'Shea et al.}{2004}]{2004astro.ph..3044O} 
O'Shea B.~W., Bryan G., Bordner J., Norman M.~L., Abel T., Harkness R., 
Kritsuk A., 2004, astro, arXiv:astro-ph/0403044 


\bibitem[\protect\citeauthoryear{Oemler}{1974}]{1974ApJ...194....1O} Oemler 
A., Jr., 1974, ApJ, 194, 1 


\bibitem[\protect\citeauthoryear{Ostriker}{1980}]{1980ComAp...8..177O} 
Ostriker J.~P., 1980, ComAp, 8, 177 


\bibitem[\protect\citeauthoryear{Park, Gott, 
\& Choi}{2008}]{2008ApJ...674..784P} Park C., Gott J.~R., III, Choi Y.-Y., 2008, ApJ, 674, 784 


\bibitem[\protect\citeauthoryear{Patton et al.}{2011}]{2011MNRAS.412..591P} 
Patton D.~R., Ellison S.~L., Simard L., McConnachie A.~W., Mendel J.~T., 
2011, MNRAS, 412, 591 


\bibitem[\protect\citeauthoryear{Perez et al.}{2009}]{2009MNRAS.399.1157P} 
Perez J., Tissera P., Padilla N., Alonso M.~S., Lambas D.~G., 2009, MNRAS, 
399, 1157 


\bibitem[\protect\citeauthoryear{Perez et 
al.}{2006}]{2006A&A...449...23P} Perez M.~J., Tissera P.~B., Lambas D.~G., Scannapieco C., 2006a, A\&A, 449, 23 


\bibitem[\protect\citeauthoryear{Perez et 
al.}{2006}]{2006A&A...459..361P} Perez M.~J., Tissera P.~B., Scannapieco C., Lambas D.~G., de Rossi M.~E., 2006b, A\&A, 459, 361 


\bibitem[\protect\citeauthoryear{Poggianti et 
al.}{1999}]{1999ApJ...518..576P} Poggianti B.~M., Smail I., Dressler A., 
Couch W.~J., Barger A.~J., Butcher H., Ellis R.~S., Oemler A., Jr., 1999, 
ApJ, 518, 576 


\bibitem[\protect\citeauthoryear{Rampazzo et 
al.}{2005}]{2005MNRAS.356.1177R} Rampazzo R., Plana H., Amram P., Bagarotto 
S., Boulesteix J., Rosado M., 2005, MNRAS, 356, 1177 


\bibitem[\protect\citeauthoryear{Rojas et al.}{2005}]{2005ApJ...624..571R} 
Rojas R.~R., Vogeley M.~S., Hoyle F., Brinkmann J., 2005, ApJ, 624, 571 


\bibitem[\protect\citeauthoryear{Rojas et al.}{2004}]{2004ApJ...617...50R} 
Rojas R.~R., Vogeley M.~S., Hoyle F., Brinkmann J., 2004, ApJ, 617, 50 


\bibitem[\protect\citeauthoryear{Ruhland et 
al.}{2009}]{2009ApJ...695.1058R} Ruhland C., Bell E.~F., H{\"a}u{\ss}ler 
B., Taylor E.~N., Barden M., McIntosh D.~H., 2009, ApJ, 695, 1058 


\bibitem[\protect\citeauthoryear{Schweizer 
\& Seitzer}{1992}]{1992AJ....104.1039S} Schweizer F., Seitzer P., 1992, AJ, 104, 1039 


\bibitem[\protect\citeauthoryear{Solanes et 
al.}{2001}]{2001ApJ...548...97S} Solanes J.~M., Manrique A., 
Garc{\'{\i}}a-G{\'o}mez C., Gonz{\'a}lez-Casado G., Giovanelli R., Haynes 
M.~P., 2001, ApJ, 548, 97 


\bibitem[\protect\citeauthoryear{Springel et 
al.}{2005}]{2005Natur.435..629S} Springel V., et al., 2005, Natur, 435, 629 


\bibitem[\protect\citeauthoryear{Szomoru et 
al.}{1996}]{1996AJ....111.2150S} Szomoru A., van Gorkom J.~H., Gregg M.~D., 
Strauss M.~A., 1996, AJ, 111, 2150 


\bibitem[\protect\citeauthoryear{Tonnesen, Bryan, 
\& van Gorkom}{2007}]{2007ApJ...671.1434T} Tonnesen S., Bryan G.~L., van Gorkom J.~H., 2007, ApJ, 671, 1434 


\bibitem[\protect\citeauthoryear{Visvanathan 
\& Sandage}{1977}]{1977ApJ...216..214V} Visvanathan N., Sandage A., 1977, ApJ, 216, 214 


\bibitem[\protect\citeauthoryear{Weinmann et 
al.}{2006}]{2006MNRAS.366....2W} Weinmann S.~M., van den Bosch F.~C., Yang 
X., Mo H.~J., 2006, MNRAS, 366, 2 

\bibitem[\protect\citeauthoryear{Willmer et 
al.}{2006}]{2006ApJ...647..853W} Willmer C.~N.~A., et al., 2006, ApJ, 647, 
853 


\bibitem[\protect\citeauthoryear{Wong et al.}{2011}]{2011ApJ...728..119W} 
Wong K.~C., et al., 2011, ApJ, 728, 119 


\bibitem[\protect\citeauthoryear{Woods 
\& Geller}{2007}]{2007AJ....134..527W} Woods D.~F., Geller M.~J., 2007, AJ, 134, 527 


\bibitem[\protect\citeauthoryear{Zabludoff}{2002}]{2002ASPC..257..123Z} 
Zabludoff A., 2002, ASPC, 257, 123 


\bibitem[\protect\citeauthoryear{Zabludoff 
\& Mulchaey}{1998}]{1998ApJ...498L...5Z} Zabludoff A.~I., Mulchaey J.~S., 1998, ApJ, 498, L5 


\bibitem[\protect\citeauthoryear{Zabludoff et 
al.}{1996}]{1996ApJ...466..104Z} Zabludoff A.~I., Zaritsky D., Lin H., 
Tucker D., Hashimoto Y., Shectman S.~A., Oemler A., Kirshner R.~P., 1996, 
ApJ, 466, 104 


\end{thebibliography}
\end{document}